\documentclass[aps,preprint,superscriptaddress, showkeys]{revtex4}
\usepackage{graphicx}
\usepackage{amsmath}
\usepackage{hyperref}
\hypersetup{colorlinks=true, citecolor=blue, urlcolor=blue, linkcolor=blue}

\usepackage{braket}
\usepackage{float}
\begin{document}

\renewcommand{\topfraction}{.85}
\renewcommand{\bottomfraction}{.7}
\renewcommand{\textfraction}{.15}
\renewcommand{\floatpagefraction}{.66}
\renewcommand{\dbltopfraction}{.66}
\renewcommand{\dblfloatpagefraction}{.66}
\setcounter{topnumber}{9}
\setcounter{bottomnumber}{9}
\setcounter{totalnumber}{20}
\setcounter{dbltopnumber}{9}

\title{Wigner crystallization in topological flat bands}
\author{B\l a\.{z}ej Jaworowski}
\email{blazej.jaworowski@pwr.edu.pl}
\affiliation{Department of Theoretical Physics, Faculty of Fundamental Problems of Technology, Wroc\l{}aw University of Science and Technology, 50-370 Wroc\l{}aw, Poland}
\author{Alev Devrim G\"{u}\c{c}l\"{u}}
\affiliation{Department of Physics, Izmir Institute of Technology, IZTECH, TR35430, Izmir, Turkey}
\author{Piotr Kaczmarkiewicz}
\affiliation{Department of Theoretical Physics, Faculty of Fundamental Problems of Technology, Wroc\l{}aw University of Science and Technology, 50-370 Wroc\l{}aw, Poland}
\author{Micha\l~Kupczy\'{n}ski}
\affiliation{Department of Theoretical Physics, Faculty of Fundamental Problems of Technology, Wroc\l{}aw University of Science and Technology, 50-370 Wroc\l{}aw, Poland}
\author{Pawe\l~Potasz}
\affiliation{Department of Theoretical Physics, Faculty of Fundamental Problems of Technology, Wroc\l{}aw University of Science and Technology, 50-370 Wroc\l{}aw, Poland}
\author{Arkadiusz W\'{o}js}
\affiliation{Department of Theoretical Physics, Faculty of Fundamental Problems of Technology, Wroc\l{}aw University of Science and Technology, 50-370 Wroc\l{}aw, Poland}

\newcommand{\TB}{\mathrm{TB}}
\newcommand{\SC}{\mathrm{SC}}
\newcommand{\Npart}{N_\mathrm{part}}

\begin{abstract}
We study the Wigner crystallization on partially filled topological flat bands of kagome, honeycomb and checkerboard lattices. We identify the Wigner crystals by analyzing the Cartesian and angular Fourier transform of the pair correlation density of the many-body ground state obtained using exact diagonalization. The crystallization strength, measured by the magnitude of the Fourier peaks, increases with decreasing particle density. The Wigner crystallization observed by us is a robust and general phenomenon, existing in all three lattice models for a broad range of filling factors and interaction parameters. The shape of the resulting Wigner crystals is determined by the boundary conditions of the chosen plaquette. It is to a large extent independent on the underlying lattice, including its topology, and follows the behavior of classical point particles. 
\end{abstract}
\maketitle

\section{Introduction}
In recent years, the possibility of realization of the quantum Hall effect (both integer and fractional) without a net magnetic field was intensely studied on topologically nontrivial energy bands of two dimensional (2D) lattice systems \cite{Haldane}.  The nontrivial topology of a band is described by a nonzero value of an integer topological invariant named Chern number \cite{TKNN}. When a band with Chern number $C\neq 0$ is fully filled, it exhibits Hall conductivity quantized to an integer multiple of $e^2/h$, in analogy to a fully filled Landau level in integer quantum Hall effect (IQHE). Such a system is called a Chern insulator. It was proposed that topologically nontrivial bands can arise entirely without a magnetic field in presence of artificial gauge fields acting on cold atom systems \cite{cooper1999composite, jaksch2003creation}. This proposition was later achieved experimentally \cite{aidelsburger2011experimental,miyake2013realizing,aidelsburger2013realization,ChernExperiment2}. Another way to realize such bands experimentally is to combine spin-orbit interaction with ferromagnetism \cite{ChernExperiment}.

Numerical calculations using exact diagonalization and DMRG approaches have shown that topological flat bands (TFBs), i.e. bands with nonzero Chern number and small bandwidth \cite{flatkagome, Sun} can host strongly correlated phases named Fractional Chern Insulators (FCIs) \cite{Tang,Neupert, SunNature, PRX,Zoology, Hierarchy, BeyondLaughlin,MooreReadWang,Johannes,We,cincio2013characterizing,liu2013bulk}. The FCIs are lattice analogs of the fractional quantum Hall effect (FQHE) states. Adiabatic continuity between the FCIs and FQHE states was shown for $C=1$ bands \cite{AdiabaticFQHE}. For larger Chern numbers, it was found that an adiabatic connection exists between FCIs and multicomponent FQHE states with a special, color entangled, boundary condition \cite{BlochPseudopotentials}. Moreover, the FCIs can be related to the Hofstadter model-- the tight-binding model of a lattice in presence of uniform background magnetic field, which can be regarded as a discretized version of the quantum Hall system \cite{Hofstadter}. There is no fundamental physical difference between a topological flat band and a subband of the Hofstadter model thus the lattice FQHE states in the Hofstadter model can be considered as fractional Chern insulators (see Ref. \onlinecite{AdiabaticHofstadter} and the discussion in Ref. \onlinecite{FCIDiscussion}). Such states were recently observed in bilayer graphene, which can be regarded as the first experimental demonstration of FCIs \cite{FCIExperiment}. There is a number of propositions of experimental realization of FCIs without a magnetic field, including cold atom \cite{sorensen2005fractional,palmer2006high,palmer2008optical,hafezi2007fractional,moller2009composite, kapit2010exact} and solid state systems \cite{NatureTransMet, layers1, layers2}. 

At the low density limit of partially filled highly degenerate systems, liquid phases compete with the Wigner crystals (WC) \cite{Wigner,YoshiokaFukuyamaHF, MakiZotos, MaksymED, HutchinsonED, HaldaneED, ShibataYoshiokaDMRG, ZhuQMC, YiQMC, LamGirvin, muller1996phase, maksym2000molecular, reimann2000formation, DevrimDotWC1, DevrimDotWC2}. The Wigner crystallization was studied for a broad range of systems -- electrons on surface of liquid helium \cite{helium}, quantum wires \cite{wireWC,DevrimWireWC}, quantum dots \cite{muller1996phase, maksym2000molecular, reimann2000formation, DevrimDotWC1, DevrimDotWC2}, boundaries of topological insulators \cite{ziani2015fractional,beule2016correlation}, as well as lattice systems \cite{Fratini,Noda} including trivial flat bands \cite{DasSarmaWCHoneycomb} and edge states of graphene nanoribbons \cite{DevrimGrapheneWC1, DevrimGrapheneWC2}. For Landau levels, it was predicted \cite{YoshiokaFukuyamaHF, MakiZotos, MaksymED, HutchinsonED, HaldaneED, ShibataYoshiokaDMRG, ZhuQMC, YiQMC, LamGirvin} and confirmed experimentally \cite{Andrei, Kukushkin} that WCs have lower energy than FQHE states for a sufficiently low filling,   
although this depends on the type of interaction \cite{HaldanePseudopotentials,TrugmanKivelson, ShibataYoshiokaDMRG,thiebaut2015fractional}.

The subject of the Wigner crystallization in TFBs remains largely untouched in previous works. Several authors investigated the charge ordering induced by short-range interaction at high filling factors \cite{varney2010interaction,SingleParticleGrushin, KourtisTriangular, kourtis2017weyl,kourtis2014combined,LiFCIWC,kourtis2017symmetry}. Phase diagrams of various flat-band models were obtained, showing the competition between the FCI and charge-ordered ground state \cite{KourtisTriangular, kourtis2017weyl,kourtis2014combined,LiFCIWC}. Moreover, it was found that the charge ordering can coexist with topological ordering \cite{kourtis2014combined, kourtis2017symmetry}. However, contrary to the Landau levels in which the Wigner crystallization occurs at arbitrarily low fillings, the short-range nature of interaction considered in Refs \onlinecite{varney2010interaction,SingleParticleGrushin, KourtisTriangular, kourtis2017weyl,kourtis2014combined,LiFCIWC,kourtis2017symmetry} limits this effect to a certain filling factor. 

In this work, we demonstrate the Wigner crystallization of spinless particles populating TFBs, interacting via short- and long-range potentials. We follow the exact diagonalization (ED) approach from Ref. \onlinecite{HaldaneED, MaksymED, HutchinsonED, Fratini} and calculate the exact ground states of variety of finite size systems in torus geometry on kagome, honeycomb and checkerboard lattices. A periodic pattern, corresponding to the Wigner crystal, is found in the pair correlation density (PCD). We analyze it using the Cartesian and angular Fourier transform, finding that the strength of the Fourier peaks -- corresponding to the strength of the Wigner crystallization -- increases with decreasing filling factor. While there are differences in the shapes of the WC unit cells related to the range of interaction, the results are to a large extent independent of the lattice type, in consistence with a picture of interacting classical point particles in a continuous space. Finally, we compare the results for trivial and nontrivial bands of the Haldane model, showing no significant differences between them.

\section{Model and methods}

Three lattice models with nearly flat bands are considered: kagome \cite{Tang}, honeycomb (Haldane model) \cite{Haldane, Neupert} and checkerboard \cite{Neupert, Sun}, with parameters chosen such that the lowest band of all three models is topologically nontrivial and nearly flat. For each model we have $|C|=1$, where $C$ is the Chern number of the lowest band, thus the same set of FCI phases can in principle be realized at each of them. The general form of a single-particle Hamiltonian is
\begin{equation}
H_{\TB}=\sum_{i,j}t_{ij}e^{i\phi_{ij}}c^{\dagger}_{i}c_{j}+H.c.,
\end{equation}
where $c^{\dagger}_i$ ($c_i$) is the creation (annihilation) operator at site $i$, while $t_{ij}$, $\phi_{ij}$ are model-dependent parameters, explained in Appendix \ref{sapp:models}. We consider the systems of dimensions $L_1\times L_2=aN_1\times aN_2$ in a torus geometry, with $N_1$ and $N_2$ being the number of unit cells in the two directions and $a$ a lattice constant. We fill them with $\Npart$ particles and apply the density-density interaction of the form $\hat{V}=\sum_{i,j}V(r_{ij})  n_{i}n_{j}$,  where $r_{ij}$ is the shortest distance between the two atoms $i$ and $j$, with periodic boundary conditions included \cite{PhysRevB.91.075102,DevrimGrapheneWC1, DevrimGrapheneWC2}. Note also that the other treatment of interactions in strongly correlated systems have been applied, i.e.\ the Ewald summation, where a sum over all periodic repetitions is taken into account \cite{Fratini}. It is obvious that both approaches give the same results for sufficiently short interaction range, and it was also shown that periodic images give neglecting contribution for a dipolar type of interaction \cite{PhysRevB.91.075102}. Our first choice for $V(r)$ is the screened Coulomb interaction $V^{\SC}_{\alpha}(r)=\frac{\exp(-\alpha r)}{r\exp(-\alpha)}$, where $\alpha$ is a parameter describing the range of interaction. In the limit $\alpha\rightarrow \infty$ the interaction contain only nearest-neighbor terms, while for $\alpha \rightarrow 0$ it converges to unscreened $1/r$ Coulomb interaction. We consider also the logarithmic interaction defined as $V^\mathrm{Log}_{\beta}(r)=\frac{\beta-\ln(r)}{\beta}$ for $r\leq \exp(\beta)$ and $V^\mathrm{Log}_{\beta}(r)=0$ otherwise, where short range interaction corresponds to small $\beta$, while for $\beta\rightarrow \infty$ it converges to $V(r)=1$. Both kinds of interactions are normalized, with $V(r)=1$ between nearest neighbors. 

We determine the ground state using the exact diagonalization method. We consider a projection of the full Hamiltonian of the system to a subspace of the lowest band, similarly to the lowest Landau level projection in FQHE. That is, we first solve the single-particle problem, and then construct the many-particle configuration basis out of the single-particle wavefunctions belonging to the lowest band. Since the wavefunctions are labeled by the momentum $k$, and the interaction conserves the total momentum of a many-particle state, we divide the basis into corresponding subspaces and diagonalize the Hamiltonian in each of them separately. We apply the flat band approximation, i.e. we neglect the single-particle dispersion by artificially setting the single-particle energies to zero for all $k$, which is a common procedure in the research on fractional Chern insulators and topological flat bands. In such a way, the only relevant energy scale in the calculation is two-body interaction strength. However, for the approximations to be meaningful, the interaction energy scale should be larger than the band dispersion and much smaller than the energy gap. The calculations has been performed using highly parallel ED software utilizing adaptive load-balanced on-the-fly matrix-vector multiplication or Hamiltonian storage in compressed sparse blocks format  \cite{csb}, depending on available system resources, paired with ARPACK eigensolver. The configuration basis of the largest system considered in this work – a $7 \times 10$ plaquette with $\Npart=7$ has size $\sim$ $1.2 \times 10^9$  (before division into 70 momentum subspaces in this case).

\section{Results}
\subsection{Identification of the Wigner crystal}
\begin{figure}
\includegraphics[width=0.8\textwidth]{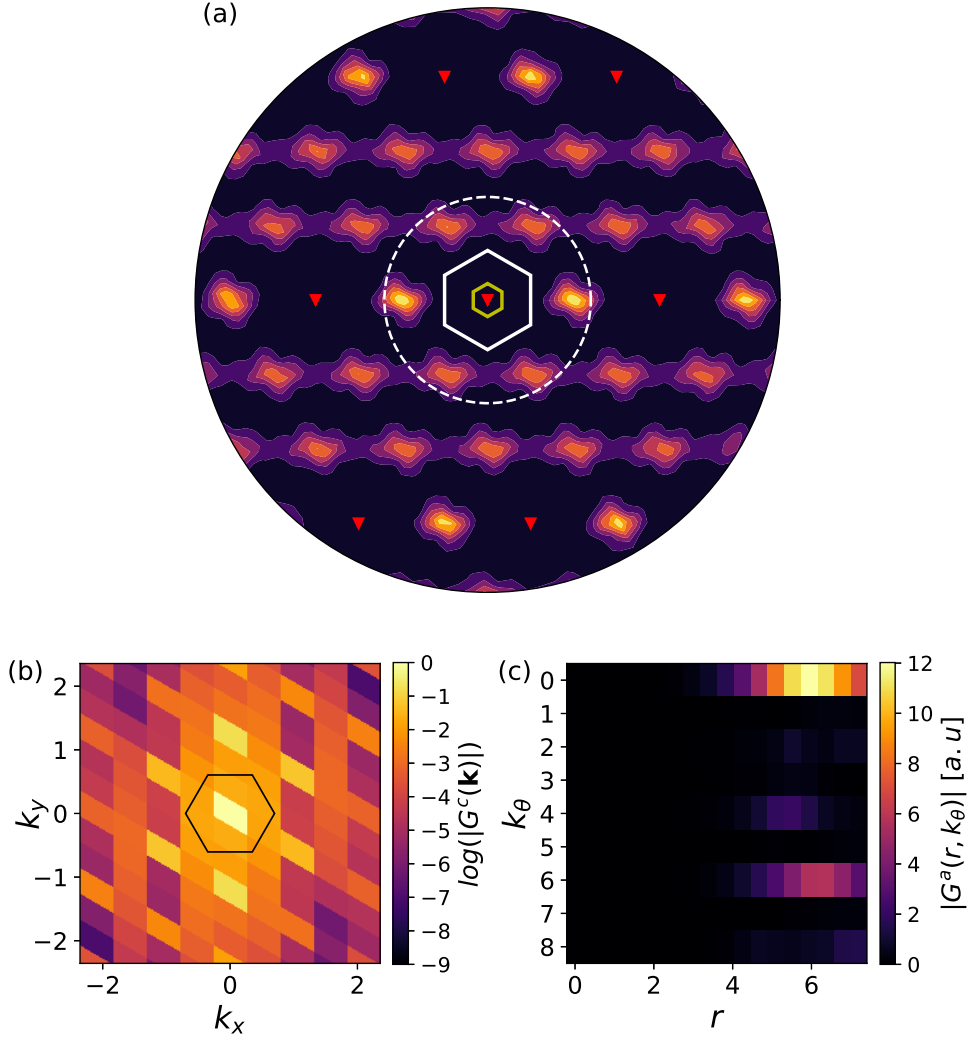}
\caption{The Wigner crystal on a $N_1 \times N_2=6 \times 9$ kagome plaquette with $\Npart=6$ particles ($\nu=1/9$ filling factor) interacting via $V^{\SC}_{0.5}$ potential. (a) The pair correlation density (PCD) of the ground state for the plaquette and its periodic images. The red triangles label the images of fixed particle. The white solid hexagon is the Wigner-Seitz unit cell of the Wigner crystal, while the smaller yellow solid hexagon is the unit cell of the underlying kagome lattice. The white dashed circle denotes the radial range used in the angular Fourier transform. (b) The Cartesian Fourier transform of the PCD. The presence of the Wigner lattice is indicated by Fourier peaks forming a hexagonal lattice described by the lattice vectors $\tilde{\mathbf{b}}_1=[\frac{\pi}{3}, -\frac{\pi}{3\sqrt{3}}]$, $\tilde{\mathbf{b}}_2=[0, \frac{2\pi}{3\sqrt{3}}]$. The scale is logarithmic and the values are normalized so that the $\mathbf{k}=[0,0]$ peak is equal to one. The black solid hexagon denotes the reciprocal space Wigner-Seitz unit cell of the WC. (c) The angular Fourier transform. The six-fold rotational symmetry of the Wigner lattice is indicated by a peak at $k_{\theta}=6$.}
\label{fig:identification}
\end{figure}

Fig. \ref{fig:identification}(a) shows the plot of the pair correlation density (PCD) $G(i,j)=\braket{\psi|c^{\dagger}_{i}c^{\dagger}_{j}c_{j}c_{i}|\psi}/\braket{\psi |c^{\dagger}_{i}c_{i}|\psi}$ of $\Npart=6$ particles with $V^{\SC}_{0.5}$ interaction on a $N_1\times N_2=6\times 9$ kagome plaquette corresponding to $\nu=\frac{\Npart}{N_1N_2}=1/9$ filling factor. The PCD is made continuous by replacing each site by a Gaussian (see the Appendix \ref{sapp:pcd}). Because our system is a torus, we repeat the plaquette to make the pattern in the PCD more visible. The red triangles mark the position of the fixed electron and its periodic images. Each maximum of the PCD corresponds to one particle forming the WC. There are $\Npart=6$ particles at each plaquette giving five maxima and one fixed particle. They are arranged in a hexagonal crystalline lattice with lattice vectors $\tilde{\mathbf{a}}_1=[6,0], \tilde{\mathbf{a}}_2=[3,3\sqrt{3}]$ and its Wigner-Seitz unit cell is marked by a white solid hexagon. As a comparison, the unit cell of the underlying kagome lattice defined by the lattice vectors $\mathbf{a}_1=[2,0], \mathbf{a}_2=[1,\sqrt{3}]$ is shown by a yellow solid hexagon, which is nine times smaller, three times in each vector direction.

The crystallization can be confirmed by looking at the plot of Cartesian Fourier transform $G^{c}$ and angular Fourier transform $G^{a}$, Fig. \ref{fig:identification}(b) and Fig. \ref{fig:identification}(c) respectively. In Fig. \ref{fig:identification}(b), there is a strong peak at zero frequency, which is the average value of the PCD. Around, there is a number of peaks arranged in a hexagonal lattice, whose lattice vectors are $\tilde{\mathbf{b}}_1=[\frac{\pi}{3}, -\frac{\pi}{3\sqrt{3}}]$, $\tilde{\mathbf{b}}_2=[0, \frac{2\pi}{3\sqrt{3}}]$, reciprocal lattice vectors to $\tilde{\mathbf{a}}_1$ and $\tilde{\mathbf{a}}_2$, in agreement with the pattern shown in Fig. \ref{fig:identification}(a). The peaks further away from the origin are weaker because the particles are not perfectly localized (see the Appendix \ref{sapp:cartesianFT} for a detailed explanation). The shape of the Wigner crystal is also probed using the angular Fourier transform in Fig. \ref{fig:identification}(c). The $k_{\theta}=0$ component is related to the value of the PCD averaged over the full angle. It is zero at $r=0$, then it increases and reaches a maximum at $r=L_1/2$ corresponding to the distance between the fixed electron and six nearest particles. Moreover, at this radius  we also see a clear component at $k_\theta=6$ as a result of a six-fold rotational symmetry of the Wigner crystal. The range of the plot in the radial direction is $r\in [0,r_0]$, where $r_0=0.6 \max(L_1,L_2)$, marked with a white dashed circle in Figure \ref{fig:identification}, to avoid the artifacts arising from the periodic images of the fixed particle. We note that the angular Fourier transform does not always look as clear as in this case. Usually the WC will be neither a perfect hexagon nor a square, hence we would obtain several peaks at frequencies $k_{\theta}=2,4,6$ or higher, possibly at different $r$ values (see the Appendix \ref{sapp:angularFT}).  Nevertheless, the highest Fourier peak will correspond to the closest symmetry.

\subsection{Wigner crystals on kagome lattice}

\begin{figure}
\includegraphics[width=0.9\textwidth]{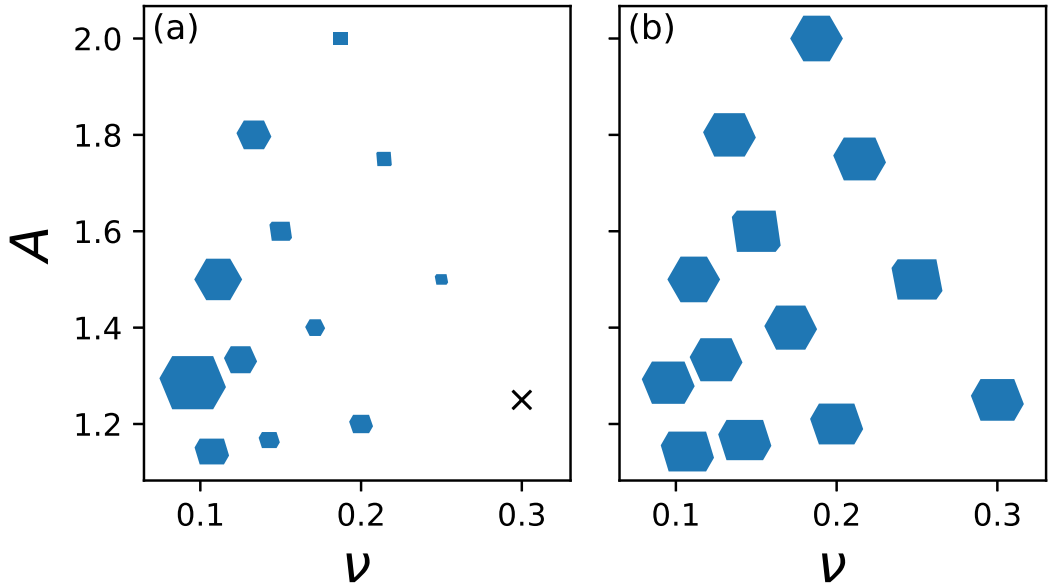}
\caption{Wigner crystallization phase diagrams for systems with $\Npart=6$ particles with $V^{\SC}_{0.5}$ interaction: (a) the ED results, (b) classical predictions. Vertical axis corresponds to the aspect ratio $A$ of plaquette, the horizontal one to the filling factor $\nu$. The shapes are the Wigner-Seitz cells of the Wigner crystal. In (a), their sizes denote the strength of the Wigner crystallization $S$. The cross denotes a liquid phase with $S$ being too small to be visible.
}
\label{fig:intrange}
\end{figure}

We move to investigate plaquettes of different size and shape. Fig. \ref{fig:intrange} (a) compares the shape of the Wigner crystal unit cells on different plaquettes of kagome lattice with screened Coulomb interaction with $\alpha=0.5$ (relatively short range interaction). We call this kind of plot a phase diagram. It contains data from a number of plaquettes with sizes from $N_1\times N_2=4\times 5$ to $N_1\times N_2=7\times 9$, each populated with $\Npart=6$ particles. Their positions on the plot denote their filling factor $\nu=\frac{\Npart}{N_1N_2}$ (horizontal axis) and aspect ratio $A=\frac{N_2}{N_1}$ (vertical axis). The blue shapes are the Wigner-Seitz cells of the Wigner crystal. The $N_1\times N_2=6\times 9$ plaquette described in the previous Subsection is situated at $\nu\approx 0.11, A=1.5$. It can be recognized by a perfectly hexagonal unit cell, although here it is rotated by 90 degrees with respect to Fig. \ref{fig:identification}. Our goal is to show the general information on the shape of the WCs. The Wigner lattices which are rotated, scaled or reflected with respect to each other are treated as the same type of WC and hence they would be indistinguishable in this plot. The size of the blue shapes denotes the strength of crystallization $S$, which we define as the product of Fourier peaks $G^c$ at two wave vectors $\tilde{\mathbf{b}}^{(i)}_1, \tilde{\mathbf{b}}^{(i)}_2$ characterizing the Wigner crystal. More precisely, a maximum value is used
$$S=\max(\{G^c(\tilde{\mathbf{b}}^{(1)}_1)G^c(\tilde{\mathbf{b}}^{(1)}_2), ..., G^c(\tilde{\mathbf{b}}^{(N_{\mathrm{W}})}_1)G^c(\tilde{\mathbf{b}}^{(N_{\mathrm{W}})}_2\}),$$ 
where the superscript index $i$ runs over $N_{\mathrm{W}}$  possible Wigner lattices (see the Appendix \ref{sapp:cartesianFT}). 

In Fig. \ref{fig:intrange}(a) it can be seen that the strength of crystallization increases with decreasing filling factor. On the smallest plaquette, $N_1\times N_2=4 \times 5$ ($\nu=0.3$ and $A\approx 1.25$), we observe a state with nearly uniform PCD, which we interpret as a liquid. On the largest plaquette considered in this phase diagram, $N_1\times N_2=7\times 9$ ($\nu\approx 0.095$ and $A\approx 1.28$), the Wigner crystal is the strongest. We do not observe clear liquid-crystal threshold filling factor but this can be related to finite size effects that will be discussed later.  The strength of the crystallization depends on the aspect ratio. The WC for $N_1\times N_2=6\times 9$ plaquette ($\nu \approx 0.111$, $A=1.5$) is stronger than the one on $N_1\times N_2=7\times 8$ plaquette ($\nu=0.107$, $A=1.14$) although the filling factor of these two is similar. A possible origin of such a dependence is the preference for the hexagonal WC. The perfectly hexagonal unit cell is allowed by the boundary conditions on plaquettes with $A=1.5$, for example the $N_1\times N_2=6\times 9$ one. Indeed, this plaquette has a second strongest WC, hence we can interpret the plot as if this aspect ratio was optimal, i.e.\ yielding the highest $S$ for fixed $\nu$. Although the $N_1\times N_2=7\times 9$ plaquette with $A=1.28$ yields a stronger WC, this may be attributed to the general trend of $S$ increasing with the decrease of filling factor.

Fig. \ref{fig:intrange}(b) shows the predictions of the WC shape from minimization of the classical energies of point like particles with short range interaction $V^{\SC}_{0.5}$ by comparing all the Wigner crystals allowed by the boundary conditions. The details of the procedure are described in the Appendix \ref{app:classical}. There is a good agreement between the resulting WC shapes and the ones obtained from ED, shown in \ref{fig:intrange}(a). We note that in the case of $L_1=L_2$ the ground state of the classical model is degenerate. If the degeneracy exists also on the ED level, the Wigner crystallization would not be detected using the product of Fourier peaks. Hence, we decided to exclude the $L_1=L_2$ plaquettes from the phase diagram and analyze them separately in the Appendix \ref{app:degeneracy}.

When we increase the range of the interaction, the strongest WCs deviate from the hexagonal shape. Similar effect is seen also for the logarithmic interaction. For both short- and longer-range $V^\mathrm{Log}$ we get a good match between classical and ED results. However, for $V^{\SC}$ the agreement deteriorates when the screening is decreased. Nevertheless, the shape of the strongest WCs is still the same as predicted classically (see the Appendix \ref{sapp:hexbrav}). 
 
\subsection{Wigner crystal on other lattices}
\begin{figure}
\includegraphics[width=0.9\textwidth]{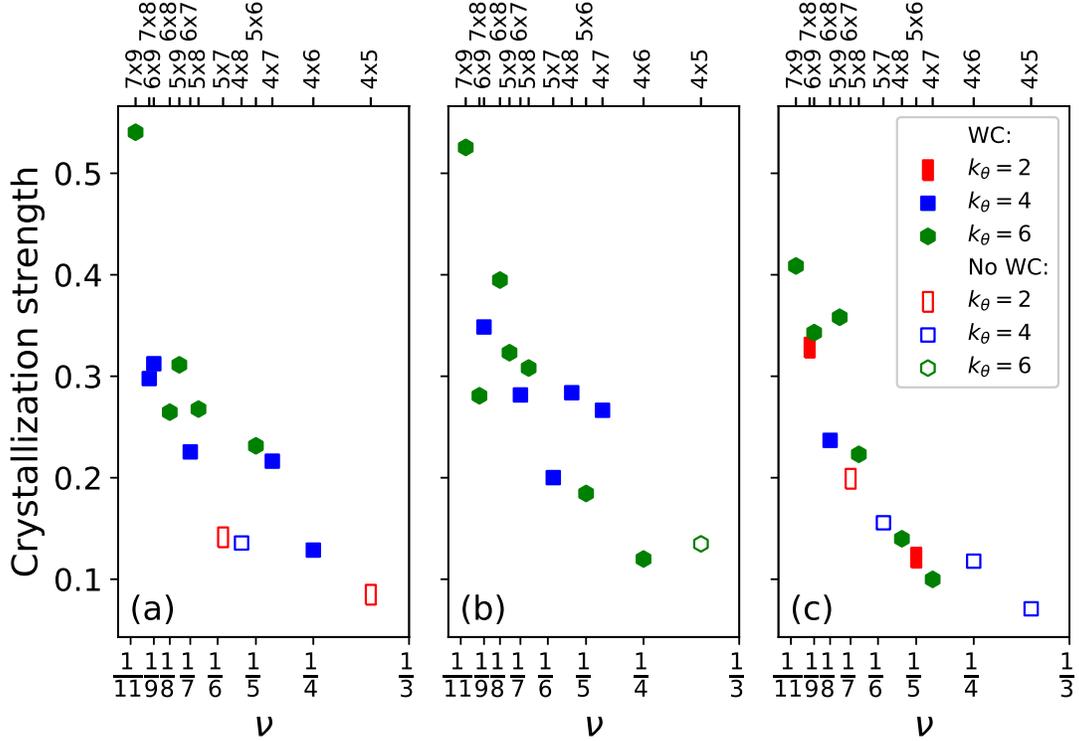}
\caption{Comparison of angular Fourier components for plaquettes of (a) kagome, (b) honeycomb and (c) checkerboard lattices with $V^{\SC}_{0.3}$. The angular components with frequencies $k_{\theta}=2,4,6$ were compared for each plaquette and only the highest ones were plotted, with frequency indicated by the color and shape of the point. Full and empty symbols denote the existence and nonexistence of a WC, respectively. The values are normalized using the procedure described in the text.}
\label{fig:angular}
\end{figure}

In Fig. \ref{fig:angular} we analyze the liquid -- crystal transition on all three lattices: (a) kagome, (b) honeycomb, (c) checkerboard. The crystallization strength is now measured by the angular transform by computing the Fourier components at $k_{\theta}=2,4,6$ and choosing the value of the strongest one. This value is normalized by dividing it by the maximum value of $k_{\theta}=0$ Fourier component within the range $r\in (0,r_0)$, with $r_0=0.6\min(L_1,L_2)$ as defined previously. Clearly, $k_{\theta}=4$ and $k_{\theta}=6$ corresponds to square and hexagonal WCs, $k_{\theta}=2$ describes WCs elongated in one of directions.  Since for some plaquettes we obtain a stripe ordering, which is not rotationally invariant and hence has nonzero angular Fourier components, we marked the plaquettes with empty symbols corresponding to no clear Wigner crystal. We consider interaction $V^{\SC}_{0.3}$, which has slightly larger range in comparison to previous results with $\alpha=0.5$, because on a checkerboard lattice shorter-range interactions lead to appearance of PCD patterns other than WC at low filling factors (see the Appendix \ref{sapp:checkerboard}).

Below filling factor $\nu=1/4$, WCs occur in most of the cases in all Fig. \ref{fig:angular}(a)--(c). Similarly to the results presented in Fig. \ref{fig:intrange}, there is no clear filling factor threshold leading to the appearance of crystallization. One can see that plaquettes with the same $\nu$ but different lattices may yield WCs with different symmetry. This can be observed e.g.\ for  $\nu\sim 1/5$. Nevertheless, the pattern of the crystallization strength smoothly increasing with lowering $\nu$ is similar for all three models, with the strongest hexagonal WC for the largest system on this phase diagram with $N_1\times N_2=7\times 9$. Comparing the kagome and honeycomb lattices (Fig. \ref{fig:angular}(a) and (b), respectively) is especially important, because both lattices have hexagonal Bravais lattice. The plaquettes with the same $N_1, N_2$ differ only by a scale factor $\sqrt{3}/2$, and hence classically they should yield similar WCs. Indeed, the strong WCs tend to have the same symmetry on both lattices, although there are counterexamples (e.g.\ $N_1\times N_2=7\times 8$). The results are also comparable for short-range and long-range logarithmic interactions. We observe significant differences between the WCs on both lattices only if we consider the Coulomb interaction with small screening. A more detailed description of the results for different interaction parameters is presented in Appendix \ref{sapp:hexbrav}.

We note that Wigner crystallization in a presence of kagome or honeycomb lattice (pinning arrays) was considered for vortices in a superconductor \cite{vortices1, vortices2}. These vortices behave like classical particles and significant differences in a crystallization pattern are observed between the kagome and the honeycomb lattices. However, the setup considered in Refs. \onlinecite{vortices1, vortices2} allows also the particles to locate at interstitial positions, which is not possible in our models. Additionally, they considered filling factors much larger than in our work, leading to much larger Wigner lattice constant. As the Wigner lattice constant grows, the influence of the lattice decreases, because the particle positions become less discretized. We note that this may be the reason why we do not observe significant lattice effects. However, it is important to emphasize that we investigate small system sizes, much smaller than in Refs. \onlinecite{vortices1, vortices2}, and also small number of particles, thus we do not rule out the possibility of the existence of larger differences between the lattices for larger systems. 

The WCs on the checkerboard lattice (Fig. \ref{fig:angular}(c)) differ from the ones on two other lattices. This stems from the fact that its Bravais lattice is square rather than hexagonal, hence the shape of the plaquettes is different. This results in a different set of WCs allowed by the boundary conditions. At low filling factors, hexagonal WCs are the strongest, but elongated hexagonal WCs appear also, as a nearly-regular hexagon can not be fitted in some plaquettes (for example for $N_1\times N_2=7\times 8$, $k_{\theta}=2$). At several plaquettes, we observe deformed WCs, where some of the particles are displaced from the ideal positions in the Wigner lattice. In some particular cases they can be predicted by minimizing the energy of classical particles, but in general the classical model is not sufficient to explain this effect. The WCs are stable for long range interactions, while decreasing their range leads to appearance of non-periodic patterns, named by us Wigner patterns (WP). Their emergence can be explained within the classical model (see the Appendix \ref{sapp:checkerboard}).

\subsection{Finite-size effects}
\begin{figure}
\includegraphics[width=\textwidth]{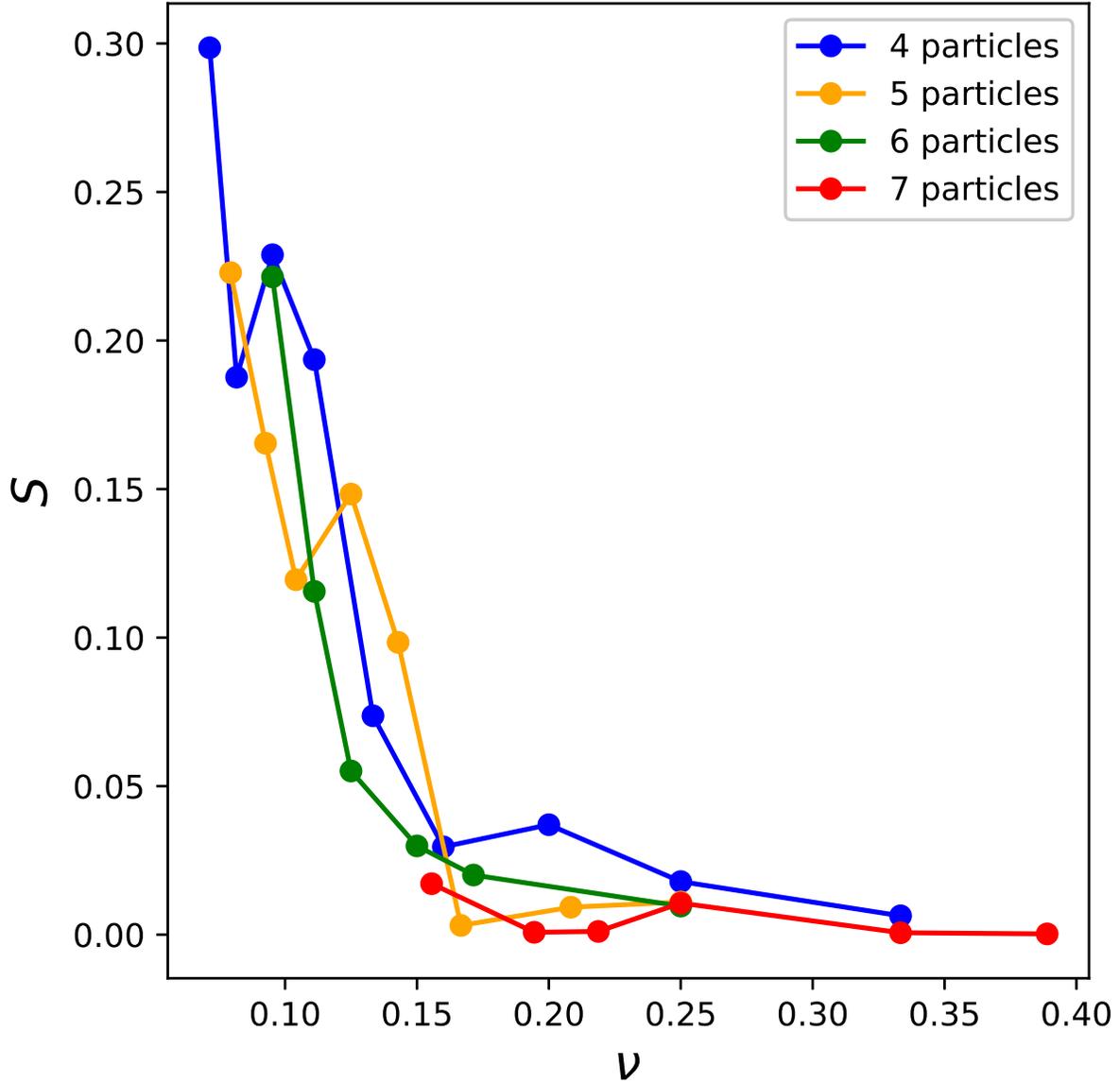}
\caption{The crystallization strength $S$, obtained from the Cartesian Fourier peaks, as a function of the filling factor, for $\Npart$ varying from 4 to 7, for the kagome lattice with $V^{SC}_{0.5}$ interaction. To minimize the effects of the aspect ratio, the plot shows only the result in a certain range of $A$: $A \in [1.0, 1.2]$ for $\Npart=4$,  $A \in [1.2, 1.6]$ for $\Npart=5$, $A \in [1.14, 1.6]$ for $\Npart=6$ and $A\in [1.4, 1.5]$ for $\Npart=7$.}
\label{fig:compare_sizes_kagome}
\end{figure}

To investigate the dependence of the Wigner crystallization on particle number, we consider systems with $\Npart$ different than 6. In Appendix \ref{app:npart}, plaquettes with $\Npart=4$, $\Npart=5$ particles are investigated. We find a good agreement between the classical model and ED results even for long range Coulomb interaction. In general, these results are consistent with the ones for $\Npart=6$ particles. It is important to note that the Wigner crystals allowed by the boundary conditions are different for every value of $\Npart$. This means that our results depend strongly on the geometric factors. For example, the optimal aspect ratio to fit a hexagonal WC with $\Npart=4$ is 1, not 1.5 as in case of $\Npart=6$. 

Now, we want to analyze liquid-WC transition regardless of the shape of WC. To find out how the Wigner crystallization is affected by the finite size effects, we compare the results for $\Npart=4,5,6$ described above and complement them also with results for $\Npart=7$. In Fig. \ref{fig:compare_sizes_kagome} we show the crystallization strength $S$, computed using the Cartesian Fourier transform, as a function of filling factor for the kagome lattice with short-range interaction $V^{\mathrm{SC}}$. Each curve corresponds to a different value of $\Npart$. To minimize the influence of the geometric factors, we show the results only for plaquettes lying within a small range of aspect ratio $A$ for which the crystallization is the strongest: $A \in [1.0, 1.2]$ for $\Npart=4$,  $A \in [1.2, 1.6]$ for $\Npart=5$, $A \in [1.14, 1.6]$ for $\Npart=6$, and we add extra results with $\Npart=7$ particles for $A \in [1.4, 1.67]$. Figure \ref{fig:compare_sizes_kagome} shows the crystallization strength $S$ on kagome lattice for the Coulomb interaction $V^{\mathrm{SC}}_{0.5}$. It can be seen that the curves corresponding to different particle numbers have a similar behavior, increasing with lowering a filling factor. The rapid increase of the crystallization strength $S$ with decreasing filling factors $\nu$ starts to occur at $\nu\approx 0.15$, i.e. close to $\nu=1/7$, although the curves for $\Npart=6,7$ are shifted towards lower filling factor with respect to curves for $\Npart=4,5$.

The shapes of the curves in Fig \ref{fig:compare_sizes_kagome} should be related to the results from Fig. \ref{fig:angular}, where crystallization occurs even for $\nu=0.25$. However, crystallization strength $S$ calculated from the multiplication of two peaks may be less sensitive to weak WC and more sensitive to strong crystallization (if the magnitude of the two peaks is roughly the same, it increases quadratically with the peak magnitude). Thus, there are weak Wigner crystals even above the rapid increase of $S$ at $\nu\approx 0.15$.

We note that the plot for $\Npart=7$ ends at plaquette $6 \times 9$, with relatively high filling factor $\nu \approx 0.13$. This is because on the plaquettes $7\times 10$ and $7 \times 11$, which are closest to $6 \times 9$ in terms of aspect ratio from all the $N_1=7$ plaquettes, we do not observe the Wigner crystallization. We interpret this result as a signature of the sensitivity of the Wigner crystal made of 7 particles to the aspect ratio of the plaquette. This may be connected with the fact that one cannot realize a nondegenerate hexagonal Wigner crystal with 7 particles.

The analysis of finite size effects for other lattices and for the long-range potential $V^{\mathrm{SC}}_{0.0}$ is presented in Appendix \ref{app:finitesize}. The behavior of the $S$ vs. $\nu$ curves is similar to what is shown in Figure \ref{fig:compare_sizes_kagome}. We note that neither in Fig. \ref{fig:compare_sizes_kagome} nor in results in Appendix \ref{app:finitesize} we do not observe the liquid-crystal transition becoming more abrupt as the number of particles increases. However, this does not necessarily mean that in the thermodynamic limit the transition will be continuous. We note that the numbers of particles investigated by us are rather small. Moreover, the behavior of the Wigner crystal depends strongly on the geometry of the sample. Thus, the reliability of the extrapolation to the infinite system is limited. Our results do not allow to determine whether the continuous nature of the transition persists in the thermodynamic limit, or is just a consequence of the small size of investigated system.

We note that the finite size effects can influence not only the profile of $S$ vs. $\nu$ curves, but also the shape of the Wigner crystals. We analyze this effect in Appendix \ref{app:finitesize}. Also, we do not rule out the possibility that there are effects which are not captured by our calculation due to the small size of plaquettes. For example, it might occur that structural changes in the Wigner crystal can happen for larger systems and that the phase diagrams of larger systems are richer than the ones we obtained. 

\subsection{Band topology}
To check how the band topology influences our results, we compared the Wigner crystallization of $\Npart=6$ on trivial and nontrivial Haldane model. We have found no significant differences between these two cases (see the Appendix \ref{app:triv}). This can be contrasted with earlier results for $\nu=1/3$ and $\nu=2/3$, where the topology is important in the description of the system, as the phase diagram contains both charge ordered and topologically ordered phases \cite{KourtisTriangular, kourtis2014combined, LiFCIWC}, however we consider lower filling factors, where FCI phases are less stable. We think that the WC-to-FCI transition can be triggered by modifying the interaction, in analogy to varying the pseudopotential parameters in FQHE.

\section{Summary and conclusions}
In summary, we have shown that the Wigner crystallization occurs in topological flat bands for low particle densities in all three considered lattice models and with a variety of interaction parameters determining the interaction range. The Wigner crystallization strength increases smoothly with decreasing filling factor. In our finite-size calculation, the WC shape depends strongly on the size and shape of the plaquette and the number of particles, which determine WCs allowed by the boundary conditions. The WC shapes were to a large extent independent on the details of the lattice type and followed the predictions made by comparing the classical energies of crystals of point-like particles in a continuous space. The underlying lattice is important only for certain aspects of the Wigner crystallization, such as the phase diagram for unscreened Coulomb interaction and the WC deformations on checkerboard lattice. 

We do not observe a sharp threshold below which the crystallization starts, but this can be related to finite size effects, which can not be eliminated from calculations presented in this work. However, we can summarize that in all our systems with various lattice models, particle numbers and interaction types, the strong Wigner crystals always occur at the lowest filling factors. The rapid increase of crystallization strength with decreasing filling factor starts at filling $\nu=1/7$ or higher. Also, we note that the agreement between the classical model and ED results exists despite the finite size effects. If it persists in the thermodynamic limit, the resulting Wigner lattice for an infinite system with an interaction $V^{\mathrm{SC}}$ will be hexagonal \cite{bonsall1977some, peeters1987wigner}. 

We have found no significant influence of band topology on the formation of the Wigner crystals. This is in contrast to earlier results obtained for $\nu=1/3$ and $\nu=2/3$ with short-range interaction and is consistent with the observation that the long-range interaction usually destroys the FCIs. 

\appendix

\section{Model and methods -- details}\label{app:methods}
\subsection{Chern insulator flat band models}\label{sapp:models}
\begin{figure*}
\includegraphics[width=\textwidth]{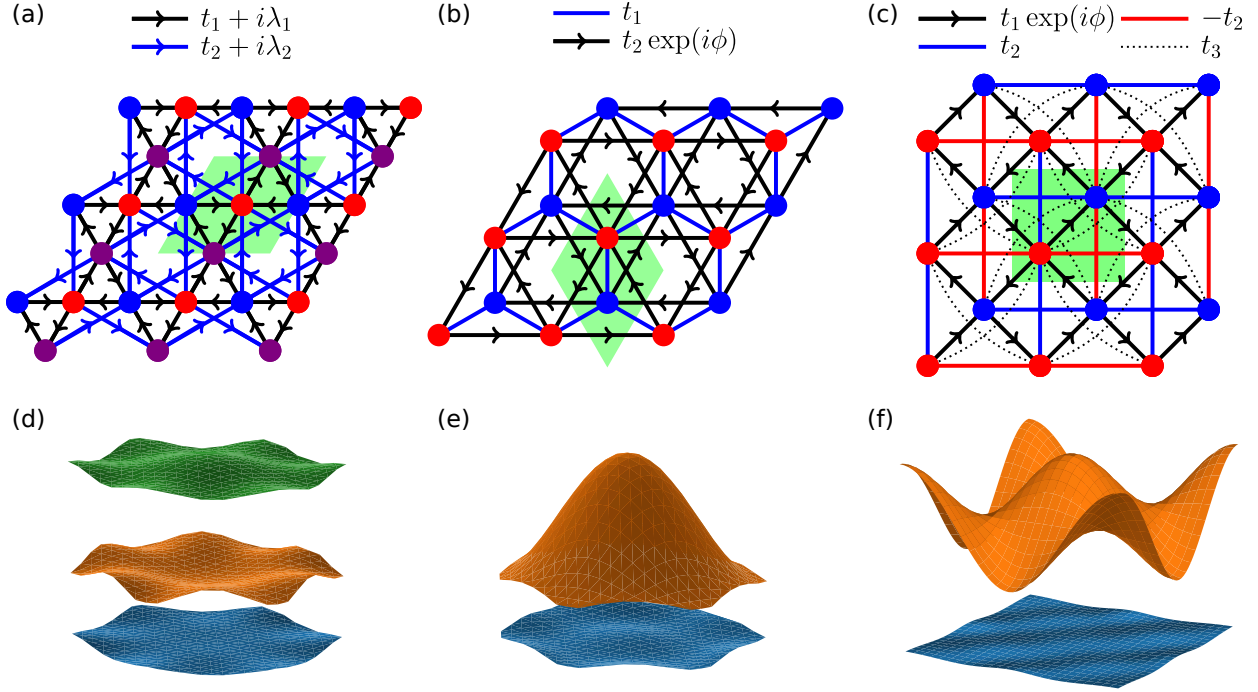}
\caption{The lattice models used in our work: (a),(d) kagome lattice, (b),(e) honeycomb lattice (Haldane model), (c),(f) checkerboard lattice. The hopping parameters are shown in the upper row, while the lower contains the band structures. The complex hoppings correspond to a particle moving in the direction denoted by arrows. Green parallelograms denote the unit cells.}
\label{fig:lattices}
\end{figure*}

The single-particle Hamiltonian of the kagome model \cite{Tang} reads
\begin{equation}
H_{\mathrm{kag}}=\sum_{\braket{i,j}}(t_1+i\nu_{ij}\lambda_1)c^{\dagger}_{i}c_{j}+ \sum_{\braket{\braket{i,j}}}(t_2+i\lambda_2\nu_{ij})c^{\dagger}_{i}c_{j},
\end{equation}
where $c^{\dagger}_{i} (c_{i})$ is a creation (annihilation) operator at site $i$, $\braket{}$, $\braket{\braket{}}$,  denote the first and the second neighbors, respectively, $t_1$ and $t_2$ are the real parts of first and second neighbour hoppings, $\lambda_1$, $\lambda_2$ are their imaginary parts, and $\nu_{ij}=\pm 1$ depending on the direction of hopping (see Fig. \ref{fig:lattices}(a)).  

The Hamiltonian of the Haldane model \cite{Haldane} is 
\begin{equation}
H_{\mathrm{hc}}=t_1\sum_{\braket{ij}}c^{\dagger}_{i}c_{j} +t_2\sum_{\braket{\braket{i,j}}}e^{i\phi_{ij}}c^{\dagger}_{i}c_{j}+\sum_{i}\epsilon_i c^{\dagger}_{i}c_{i},
\end{equation}
where $t_1$ and $t_2$ are magnitudes of the first and second neighbor hoppings, respectively, $\phi_{ij}=\pm\phi$ is a complex phase with a sign depending on the direction of hoppings, shown in Fig. \ref{fig:lattices}(b). $\epsilon_i\pm \epsilon$ is the staggered onsite potential, $+\epsilon$ on red sublattice and $-\epsilon$ on the blue one.

The checkerboard model \cite{Sun,Tang} is described by the Hamiltonian 
\begin{equation}
H_{\mathrm{cb}}=t_1\sum_{\braket{i,j}}e^{i\phi_{ij}}c^{\dagger}_{i}c_{i}+
\sum_{\braket{\braket{i,j}}}t'_{ij} c^{\dagger}_{i}c_{j}
+t_3\sum_{\braket{\braket{\braket{i,j}}}}c^{\dagger}_{i}c_{j},
\end{equation}
where $t'_{ij}=\pm t_2$ depends on the sublattice and the direction of the hopping, as indicated in Fig. \ref{fig:lattices}(c), $t_\alpha$, with $\alpha=1,2,3$ denoting the absolute values of $\alpha$th-neighbor hopping. The nearest-neighbor hopping contains a complex term with a phase $\phi_{ij}=\pm\phi$, where the sign corresponds to clockwise or counterclockwise direction of the hopping. 

In all three models, the parameters can be tuned so that the lowest band is topologically nontrivial with $|C|=1$ and nearly flat \cite{Sun,Tang,Neupert}. In the course of this work, we use for kagome model $t_1=-1$, $t_2=0.3$, $\lambda_1=0.6$, $\lambda_2=0$, $t_1=1$, for honeycomb model $t_2=\frac{\sqrt{43}}{12\sqrt{3}}$, $\phi=\arccos \left (3\sqrt{\frac{3}{43}} \right )$, $\epsilon=0$, $t_1=1$ and for checkerboard model $t_1=\frac{1}{2+\sqrt{2}}$, $t_2=\frac{1}{2+2\sqrt{2}}$, $\phi=\pi/4$. The corresponding band structures are plotted in Fig. \ref{fig:lattices}(d)-(f). We also investigate the trivial version of the Haldane model, with the lowest band topologically trivial, with parameters $t_1=1$, $t_2=\frac{\sqrt{43}}{12\sqrt{3}}$, $\phi=\arccos \left (3\sqrt{\frac{3}{43}} \right )$ and $\epsilon=0.15$.

We consider finite-size systems in torus geometry, i.e. we investigate finite plaquettes of $N_1 \times N_2$ unit cells with periodic boundary conditions. The lattice is defined by lattice vectors $\mathbf{a}_1, \mathbf{a}_2$, so the dimensions of the plaquette are $L_{1,2}=|\mathbf{a}_{1,2}|N_{1,2}$. For all the lattices we consider, we have $|\mathbf{a}_1|=|\mathbf{a}_2|=a$. The scale of $|\mathbf{a}_{1,2}|$ is determined by the distance $d$ between the nearest neighbor sites, which we fix to be $d=1$.

\subsection{Pair correlation density}\label{sapp:pcd}
Having obtained the ground state $\ket{\psi}$, we calculate the pair correlation density (PCD)
\begin{equation}
G(i,j)=\frac{\braket{\psi|c^{\dagger}_{i}c^{\dagger}_{j}c_{j}c_{i}|\psi}}{\braket{\psi |c^{\dagger}_{i}c_{i}|\psi}},
\end{equation}
defined in the discrete basis of sites, describing probability of finding a particle at site $j$ assuming that there is a fixed particle at site $i$. We make it continuous by replacing every site by a Gaussian,
\begin{equation}
G_{i}(\mathbf{r})=\sum_{j=1}^{N} G(i,j)\frac{1}{\sigma\sqrt{2\pi}}\exp\left(-\frac{|\mathbf{r}-\mathbf{r_{j}}|}{2\sigma}\right),
\end{equation}
where $\mathbf{r}$ is the vector connecting atom $i$ and a given point in space, i.e.\ we take the site $i$ as the origin of our coordinate system, and $\sigma$ is the width of the Gaussian, which we choose to be $\sigma=0.5$. The choice of starting site $i$ does not affect the results significantly, as the exact-diagonalization eigenstates are translationally invariant. To find the Wigner crystal, we discretize this function on a Cartesian or polar grid and perform the Fourier transform using the Fast Fourier Transform algorithm.

\subsection{The Cartesian Fourier transform}\label{sapp:cartesianFT}
If we choose the Cartesian grid, we perform the Fourier transform in both directions and obtain the Fourier coefficients
$$
G^c(\mathbf{k})=\iint_{P}d\mathbf{r} G_i(\mathbf{r})\exp(-i \mathbf{r}\cdot \mathbf{k}),
$$
where $P$ denotes the area of the plaquette, and $\mathbf{k}$ is the wave vector. Because the system is periodic, the $\mathbf{k}$ vectors can have only discrete values $\mathbf{k}=\frac{p}{N_1}\mathbf{b}_1+\frac{q}{N_2}\mathbf{b}_2$, with $\mathbf{b}_{1,2}$ being the reciprocal lattice vectors corresponding to the real-space lattice defined by $\mathbf{a}_{1,2}$, and $p,q$ being arbitrary integers.

The Wigner crystal is defined by lattice vectors $\tilde{\mathbf{a}}_{1,2}$. Because our system is a finite-size torus, only a subset of $\tilde{\mathbf{a}}_{1,2}$ vectors is allowed by the boundary conditions. Moreover, since we fix the number of particles $\Npart$, the number of PCD maxima within the plaquette should be equal to $\Npart-1$. Otherwise, the state is not a Wigner crystal but another charge ordering. Ideally, the Wigner crystal would consist of point particles arranged in a lattice, with PCD

\begin{equation}
\label{eq:deltas}
G_{\mathrm{I}}(\mathbf{r})\sim G_0(\mathbf{r})-\delta(\mathbf{r}),
\end{equation}
where
$$
G_0(\mathbf{r})\sim \sum_{m, n}\delta(\mathbf{r}-m\tilde{\mathbf{a}}_1-n\tilde{\mathbf{a}}_2),
$$
with $m,n$ being arbitrary integers and $\delta(\mathbf{r})$ being the Dirac delta. The delta at the origin is subtracted because the fixed particle is not included in the pair correlation function.

The Fourier transform of $G_0$ would be an infinite sum of periodically arranged Dirac deltas,
$$
G^c_0(\mathbf{k})\sim \sum_{m, n=-\infty}^{\infty}\delta(\mathbf{k}-m\tilde{\mathbf{b}}_1-n\tilde{\mathbf{b}}_2),
$$
where $\tilde{\mathbf{b}}_{1,2}$ are the reciprocal lattice vectors of WC, each of them given by a pair of two integers $\tilde{p}_{i}, \tilde{q}_{i}$, $\tilde{\mathbf{b}}_{i}=\frac{\tilde{p}_i}{N_1}\mathbf{b}_1+\frac{\tilde{q}_i}{N_2}\mathbf{b}_2$. Not every choice of $\tilde{p}_{i}, \tilde{q}_{i}$ is permitted, as they should yield a correct number of PCD maxima.

The Fourier transforms we obtain in ED calculations are not as ideal as $G^c_0(\mathbf{k})$ for two reasons. First, the particles have finite spatial dimensions. This can be seen on a simple example of particles described by Gaussians of width $\sigma_\mathrm{W}$. Then, the PCD will be a convolution of $G_{\mathrm{I}}$ with a Gaussian
$$
G_{\mathrm{Gauss}}(\mathbf{r})=\int d\mathbf{r} G_\mathrm{I}(\mathbf{r})\exp \left (-\frac{r^2}{2\sigma_\mathrm{W}} \right )
$$
Usung Eq. \ref{eq:deltas}, we get
\begin{equation}
\label{eq:gaussian}
G_{\mathrm{Gauss}}(\mathbf{r})\sim G_{\mathrm{G0}}(\mathbf{r}) -\exp \left (-\frac{r^2}{2\sigma_\mathrm{W}} \right ),
\end{equation}
with
$$
G_{\mathrm{G0}}(\mathbf{r})=\int d\mathbf{r} G_0(\mathbf{r})\exp \left (-\frac{r^2}{2\sigma_\mathrm{W}} \right ).
$$
The Fourier transform of $G_{\mathrm{G0}}$ is a multiplication of $G_0(\mathbf{k})$ and a Gaussian in a momentum space
$$
G^c_{\mathrm{G0}}(\mathbf{k})=G_0(\mathbf{k})\exp \left (-\frac{\sigma_{\mathrm{W}}}{2} k^2 \right ).
$$
Therefore, the spatial delocalization makes the Fourier peaks decay with increasing distance from the origin -- an effect which is visible in Fig. 1(b) of the main text. 

Another source of distortion from the ideal periodic pattern is the fact that the fixed particle is not included in the pair correlation density. The subtracted delta in Eq. \ref{eq:deltas} and Gaussian in Eq. \ref{eq:gaussian} will give rise to additional Fourier components at $\mathbf{k}$ vectors not belonging to the reciprocal lattice of the Wigner crystal. Similar efect is observed in our numerical results. The spurious Fourier components are visible as the bright ''cloud'' around the origin in Fig. 1(b) of the main text.

We use the magnitude of the Fourier peaks as the measure of the strength of the Wigner crystallization. The WC has to be periodic in two directions, hence we should observe at least two nonzero peaks. Therefore we choose our measure to be a product of two peaks
$$
S_i=G^c(\tilde{\mathbf{b}}^{(i)}_1)G^c(\tilde{\mathbf{b}}^{(i)}_2),
$$
where $\tilde{\mathbf{b}}^{(i)}_{1,2}$ are the two reciprocal lattice vectors defining the WC of a given type indexed by $i$. If the PCD is non-periodic in at least one direction, this product will vanish. We do not know which WC will be present on which plaquette. Therefore, we first list all the possible $N_{\mathrm{W}}$ Wigner crystals and their lattice vectors. For example for $\Npart=6$ particles on kagome lattice $N_\mathrm{W}=8$. Since the dimensions $L_{1,2}$ of the plaquettes differ, these vectors will be different at each of them. Nevertheless, they will be defined by the same $(\tilde{p}, \tilde{q})$ pairs. To determine which WC is present on the plaquette, we check which pair of reciprocal lattice vectors gives the highest product $S_i$ of the Fourier components. This product is then taken as the crystallization strength $S$.

Several comments need to be made here. First, to compare the results for different plaquettes, the Fourier spectrum has to be normalized, which is done by dividing it by the $\mathbf{k}=[0,0]$ component. Secondly, the ''holes'' in the PCD corresponding to the fixed electron may introduce nonzero Fourier components at $\tilde{\mathbf{b}}^{(i)}_{1,2}$ defining the Wigner crystals even if there is in fact no WC. Indeed, some of the small unit cells in Fig. 2(a) of the main text do not correspond to WCs. However, if strong WC is present, the peaks due to WC will dominate over the spurious Fourier components, as can be seen in Fig. 1(b) of the main text. Finally, the choice of the reciprocal lattice vectors describing a given WC is to some extent arbitrary, as we can choose different unit cells. Usually there are several choices of the unit cells which have similarly strong peaks. We choose one of them arbitrarily and use this choice consistently for every plaquette (i.e. we use vectors defined by the same $\tilde{p}$ and $\tilde{q}$). Although making a different choice may affect the value of $S$ for some weak Wigner crystals, it would not change the general picture.

\subsection{The angular Fourier transform}\label{sapp:angularFT}
\begin{figure}
\includegraphics[width=0.9\textwidth]{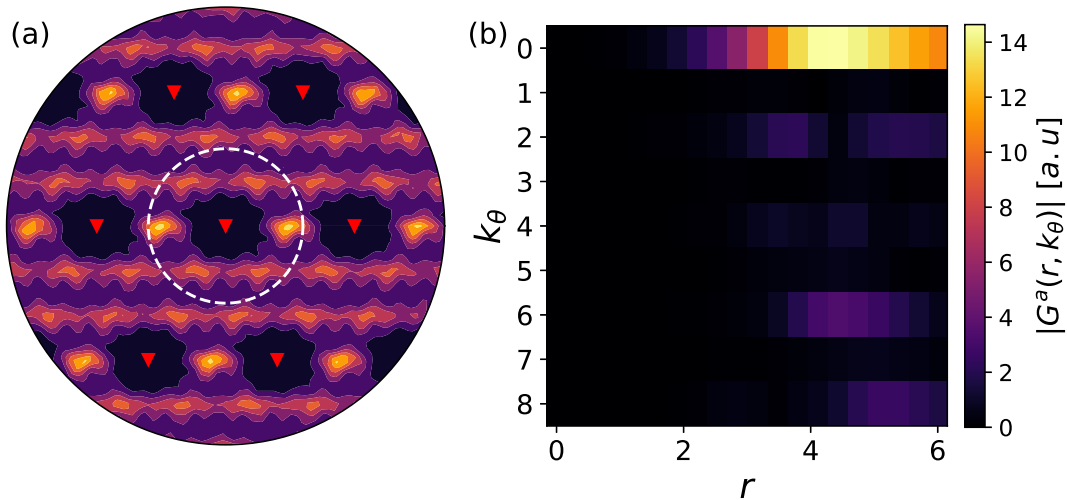}
\caption{The pair correlation density (a) and its angular Fourier transform (b) for $N_1 \times N_2=5\times 6$ plaquette with $\Npart=6$ particles with $V^{\SC}_{0.3}$ interaction. There are several Fourier components, each exhibiting a maximum at different radius. The $r$ range in (b) corresponds to the white dashed circle in (a).}
\label{fig:nonideal}
\end{figure}
Another choice of discretization of $G(\mathbf{r})$ is the polar grid. Then, the Fourier transform is taken only along the angular direction, and the Fourier components are given by
$$
G^a(r, k_{\theta})=\int_{0}^{2\pi} d\theta G_i(r, \theta)\exp(-i \theta k_\theta),
$$
where $k_\theta$ is the angular frequency. The $k_\theta=0$ component is related to the average PCD at radius $r$, while all the others allow to distinguish the lattice symmetry. In the case of a nearly-hexagonal or nearly-square WC, the Fourier transform will contain a strong component at $k_{\theta}=6$ or $k_{\theta}=4$, respectively. As noted in the main text, it would occur at the radius equal to the distance between the first particle and the six or four nearest particles. At this radius, the zeroth component would exhibit its first maximum.

The transform is not meaningful at large $r$. The ''holes'' in PCD due to the presence of periodic images of fixed electron introduce at least 2-fold rotational symmetry and therefore nonzero Fourier component even for perfectly isotropic liquid state. Therefore, we have to introduce a cutoff $r_0$. Strictly speaking, the influence of the periodic images of fixed electron starts at half the distance to the closest of them, i.e. $r=0.5 \max(L_1,L_2)$. However, we note that often a particle is located at this distance or even further, therefore the cutoff has to be slightly larger. We choose $r_0=0.6 \max(L_1,L_2)$.

Also, we note that the angular Fourier transform does not always look as clear as in Fig. \ref{fig:identification}(c) of the main text (see Fig. \ref{fig:nonideal}). If the Wigner lattice is not close to neither hexagonal nor square symmetry, we would obtain several strong Fourier components at even frequencies (the odd components will vanish at least approximately because all the possible Wigner lattices have a 2-fold rotational symmetry). Moreover, if $|\tilde{\mathbf{a}}_1|\neq|\tilde{\mathbf{a}}_2|$ the maxima of different Fourier components may occur at different radii, hence the PCD may exhibit different symmetries at different $r$ (see Fig. \ref{fig:nonideal}). To determine which rotational symmetry (2, 4- or 6-fold) is the closest one, we compute the maximal value of Fourier components with $k_{\theta}=2,4,6$ in the range $[0, r_0]$. The $k_{\theta}$ at which the value is the highest indicates the symmetry of WC. We use this value as an alternative measure of crystallization strength $\tilde{S}$. However, since the magnitude of Fourier components depends on the mean particle density, we normalize it by dividing by the maximal value of $k_{\theta}=0$ component in the range $[0, r_0]$.

As we noted in the main text, $\tilde{S}$ can be nonzero even if the system is not a WC (for example a stripe phase would also have 2-fold rotational symmetry). Therefore we have to select the WCs first, can be done visually by looking at the PCD plot, or comparing with the results for Cartesian Fourier transform.

\section{Classical model}\label{app:classical}
We compare the shapes of WCs obtained from the exact diagonalization calculation to predictions made using a simple classical model. The classical energy of a set of point particles is given by
$$
E=\frac{1}{2}\sum_{i\neq j}V(r_{ij})
$$
where the indices $i$ and $j$ run over all the particles, and $r_{ij}$ is the shortest interparticle distance on the torus. The classical prediction of the WC shape is found by calculating this energy for every Wigner lattice allowed by the boundary conditions, and choosing the one in which $E$ is minimal. We do not take the underlying lattice into account, i.e. the particle position is not restricted to lattice sites, and is determined only by the Wigner lattice.

Such a model allows also for introduction of patterns other than the perfect crystal. We will consider several such shapes, parameterized by a single number $\delta$ (e.g. the displacement of some particle from ideal crystal positions). For each pattern like this, the energy is minimized with respect to $\delta$ and then compared with the energies of other patterns and WCs.

We note that for the logarithmic interaction, the particles may not interact classically if $\beta$ is too small. Then the classical model may have several zero-energy ground states. However, the interaction may still exist at the quantum level, possibly because the particles are not perfectly localized, and their positions are restricted to lattice sites. For example, for $\Npart=6$ particles on kagome lattice we have a degenerate classical ground state at $\beta<1.82$, although the ED calculations yield a nondegenerate WC even when $\beta\sim 1.4$. Because of this effect, the exact diagonalization results cannot be compared to classical predictions for certain values of $\beta$. Such a problem is not present in screened Coulomb interaction, whose exponential tail always lifts the degeneracy.

\section{Degeneracy}\label{app:degeneracy}
\begin{figure}
\includegraphics[width=0.9\textwidth]{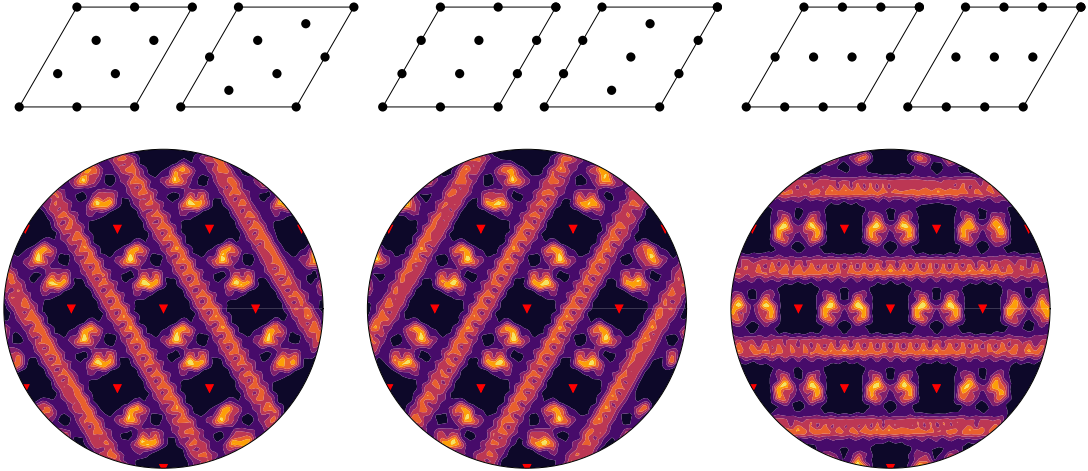}
\caption{The degenerate ground states of $N_1 \times N_2=7\times 7$ honeycomb plaquette with $\Npart=6$ particles with $V^{\SC}_{0.1}$. In the lower row PCDs are plotted. There are six degenerate ground states in total, but pairs of them have similar PCD so we plot only one state of each pair. Each of these patterns can be thought of as a superposition of two Wigner lattices, drawn schematically in the upper row.}
\label{fig:degeneracy}
\end{figure}

The plaquettes with aspect ratio $A=\frac{N_2}{N_1}=1$ were omitted in our analyses of $\Npart=6$ case. This is because the ground state will always be degenerate. For example, for the plaquettes with hexagonal Bravais lattice, the $N_{\mathrm{W}}=8$ possible Wigner crystals can be divided into two sets of WCs with the same classical energy, one consisting of six WCs, the other of two. 

Indeed, the results for $L_1=L_2$ honeycomb plaquettes obtained with certain interactions can be interpreted in such a way. There are six degenerate ground states, none of which yields a clear Wigner crystal in the pair correlation density. Instead, pairs of these states have similar, stripe-like PCD. This does not mean that the Wigner crystallization does not occur. The ground state obtained in the exact diagonalization procedure may be a superposition of degenerate groundstates. We interpret each of the stripe-like patterns as two Wigner lattices superimposed (see Fig. \ref{fig:degeneracy}). For the kagome $7\times 7$ plaquettes the ground state is also 6-fold degenerate, and the sum of their PCDs has some similarities with a superposition of all six Wigner lattices. Moreover, at smaller plaquettes we obtain a similar PCD pattern, but the ground state is 3- or 1-fold degenerate. Even if these states are indeed a superposition of Wigner crystals, we cannot measure the crystallization strength, as we would have to take into account a combination of six reciprocal-space lattices. Therefore we decided to exclude the $L_1=L_2$ plaquettes from our considerations.

\section{Different lattices}\label{app:lattices}
\subsection{Kagome and honeycomb lattices}\label{sapp:hexbrav}
\begin{figure}
\includegraphics[width=0.9\textwidth]{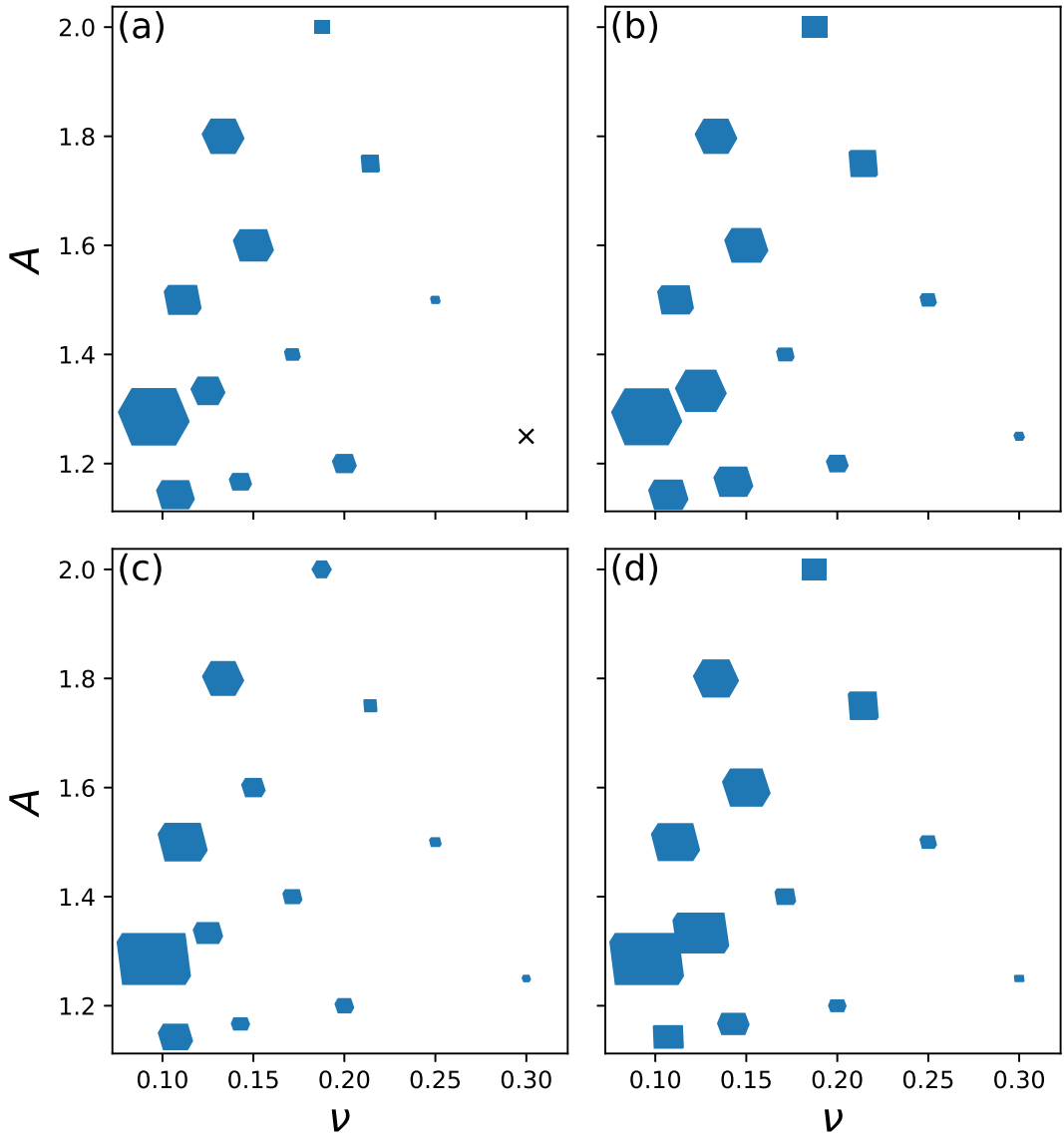}
\caption{Phase diagrams for systems with $\Npart=6$ particles. (a) Kagome lattice, $V^{\SC}_{0.3}$, (b) honeycomb lattice, $V^{\SC}_{0.4}$,(c) kagome lattice, $V^{\SC}_{0.0}$, (d) honeycomb lattice, $V^{\SC}_{0.0}$.}
\label{fig:hc}
\end{figure}
As we noted in the main text, the Wigner crystals on kagome and honeycomb plaquettes defined by the same $N_1, N_2$ are similar. The similarity is even greater if we compare different interaction ranges. Fig. \ref{fig:hc}. shows the phase diagrams for (a) kagome lattice with $\alpha=0.3$ and (b) honeycomb lattice with $\alpha=0.4$ . The Wigner crystals have exactly the same shape on corresponding plaquettes. There are differences in crystallization strength, but the strongest WCs concentrate around the maximum at $N_1\times N_2=7\times 9$ on both lattices. The difference in interaction range probably stems from the fact that honeycomb plaquettes are smaller than the kagome ones by the factor of $2/\sqrt{3}$, which is a result of the difference in the unit cell size. Hence, it is not the intersite distance scale that matters -- it is the same for both lattices -- but rather the length scale of the torus, i.e. $L_1, L_2$. For simplicity, we omitted this effect in discussions of Fig. 3 in the main text, noting that there is still a large degree of similarity between the WCs on the two lattices if we use $\alpha=0.3$ on both. 

Figure \ref{fig:hc}(c) and (d) shows the phase diagram for unscreened Coulomb interaction ($\alpha=0$) for kagome and honeycomb lattices, respectively. One can clearly see that there are more differences between these two than between (a) and (b) subfigures. In general, the similarity betwen WCs on kagome and honeycomb lattices lowers with decreasing $\alpha$. However, even if $\alpha=0$ (Fig. \ref{fig:hc}(c) and (d)), there is a considerable similarity if one limits the comparison to strong WCs only. The $N_1\times N_2=7\times 9, 6\times 9$ and $5\times 9$ plaquettes (i.e. the ones with strongest WCs in (c)) yield the same shape of WC on both lattices. Decreasing $\alpha$ leads also to deterioration of the accuracy of the classical predictions. Nevertheless, the WC shapes on the three plaquettes mentioned above are in agreement with classical results. Also, the classical model correctly predicts that increasing the range of interaction makes the WCs at lower aspect ratios deviate from the hexagonal shape, even if the exact shape of WC unit cell does not agree with ED results.

For logarithmic interaction, such a deterioration does not happen. We investigated the logarithmic interaction on kagome lattice with $\beta$ between 1.4 and 3.0 and found that at small $\beta$ the WCs seem to prefer the hexagonal shape, while for higher $\beta$ the WCs at small aspect ratios are closer to rectangular shape. This behavior is also well captured by the classical model, as long as $\beta$ is large enough that the particles interact classically. Although the details of the transition differ in classical and ED approaches, their results agree well or even perfectly at its ``end points'' at high and low $\beta$. Also, we found that the WC shapes for kagome lattice are similar to the ones for honeycomb lattice for both short ($\beta=1.3$ honeycomb, $\beta=1.4$ kagome) and longer range interaction ($\beta=3.0$ on both lattices).

\subsection{The checkerboard lattice}\label{sapp:checkerboard}
\begin{figure}
\includegraphics[width=0.9\textwidth]{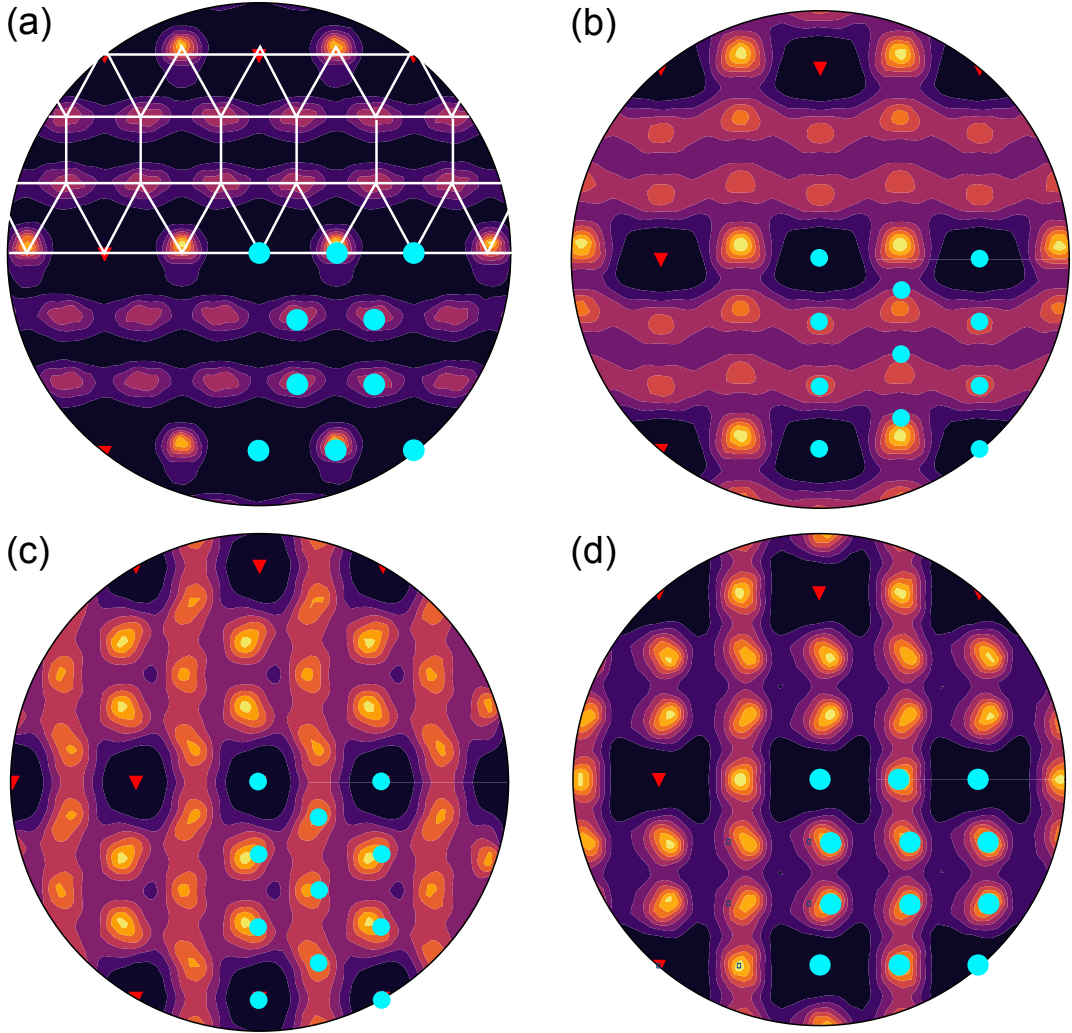}
\caption{Deviations from perfect WC on the checkerboard lattice.(a) The half-elongated Wigner pattern for a $N_1\times N_2=7\times 9$ checkerboard plaquette with $V^{\mathrm{Log}}_{1.6}$. (b), (c), (d) -- deformed WCs on (b) $N_1\times N_2=5\times 6$ plaquette with $V^{\SC}_{0.4}$, (c) $4\times 7$ plaquette with $V^{\SC}_{0.4}$, (d) $N_1\times N_2=6\times 7$  plaquette with $V^{\mathrm{Log}}_{1.6}$. The blue dots show the positions of particles obtained from the classical model.}
\label{fig:halfelongated}
\end{figure}

The checkerboard lattice is more difficult to analyze, as, in addition to Wigner crystal, liquids and stripe patterns one observes also another type of charge ordering. We call it a ``Wigner pattern'' (WP) to emphasize that it consist of well-localized particles, but exhibit no periodicity other than the periodicity of the torus. In general, many Wigner patterns are possible, but in our calculations we encounter only one. We call it ``half-elongated'', since it resembles the half-elongated triangular tiling of the plane. It consists of rows of triangles and squares, with two rows of triangles per one row of squares, with particles located in their corners (see Fig. \ref{fig:halfelongated}(a)).  Obviously, the aspect ratio of the plaquette usually does not allow the triangles and squares to be regular polygons, so the pattern is always squeezed or stretched. Also, we observe WCs in which the particles deviate from their ideal positions in the crystal lattice, but the displacement is small enough for the Wigner lattice to be identified (see Fig. \ref{fig:halfelongated}(b), (c), (d)). We will call these ``deformed WCs''.

The existence of these effects makes it more difficult or even impossible to measure the crystallization strength. The half-elongated WP cannot be described by two Fourier peaks, so we can only check visually whether it exists or not. The deformed WCs, if they are close enough to the perfect lattice, will have nonzero Fourier components corresponding to that crystal, so they may be visible using the procedure described in the main text.
We have investigated the checkerboard lattice with screened Coulomb interaction with $\alpha=0,0.1,...,1$ and logarithmic with $\beta=1.2, 1.4,...,3.0$. For sufficiently long-range interaction the Wigner crystals are common. On three plaquettes, $N_1 \times N_2=4\times 7$, $N_1 \times N_2=5\times 6$ and $N_1 \times N_2=6\times 7$, we encounter deformations, but they are small enough for the crystallization to be seen from Fourier peaks. The shapes of WCs (including the deformed ones) are the same for both interaction types on all the plaquettes. The maximum of crystallization strength occurs again at $N_1\times N_2=7\times 9$ plaquette. When the range of the interaction is decreased, more and more WPs and/or deformations start to appear, starting from low fillings and low aspect ratios. Also, for a small number of plaquettes with low fillings, we observe a charge ordering which is neither WC nor WP, as it does not correspond to six well-localized particles.

 Figure \ref{fig:halfelongated}(a) shows a comparison of the classical prediction of particle positions with the exact-diagonalization PCD for a $N_1\times N_2=7\times 9$ checkerboard plaquette with $V^{\mathrm{Log}}_{1.6}$. A good agreement between those two results is seen. In general, the classical model correctly describes the emergence of the half-elongated WP at the qualitative level. For longer-range interaction it predicts no WPs. They emerge, starting with high fillings and low aspect ratios, when the interaction range is decreased. On the quantitative level, the model does relatively well for the screened Coulomb interaction $V^{\SC}$. For example, for $\alpha=0.9$ and $\alpha=1.0$ the classical model predicts half-extended WP on five plaquettes ($N_1 \times N_2=7\times 8$, $7\times 9$, $6\times 7$, $6\times 8$, $5\times 6$), in four of which it exists also in quantum results (all the above except $N_1 \times N_2=5\times 6$). For logarithmic interaction its performance is worse. For example, for $V^{\mathrm{Log}}_{1.6}$ it predicts half-elongated WPs at four plaquettes ($N_1\times N_2=7\times 9$, $6\times 8$, $6\times 9$, $5\times 9$), while in exact diagonalization it exist on three ($N_1\times N_2=7\times 8$, $7\times 9$ and $6\times 9$), and only two are guessed correctly.
 
On the other hand, the classical model fails to describe the deformed WCs. This can be seen in Fig. \ref{fig:halfelongated}(b) and (c). In both subfigures, the classical model predicts no deformation, although they exist on the ED level. Similar behaviour is observed in the case of longer-range $V^{\mathrm{Log}}$, and $V^{\SC}$ regardless of $\alpha$. For short-range $V^{\mathrm{Log}}$, the model predicts too many deformations. Although in several cases it correctly predicts their shape (Fig. \ref{fig:halfelongated}(d)) usually the prediction is wrong. This suggests that the deformations arise rather due to the presence of the lattice. Also, we note that the deformation of the type shown in Fig.  \ref{fig:halfelongated}(b) exists only when $N_1$ is odd ($N_1\times N_2= 5 \times 6$, $5\times 8$, $5\times 9$, $7\times 9$ plaquettes) while the one in Fig.  \ref{fig:halfelongated}(c) only for even $N_1$ and odd $N_2$ ($N_1\times N_2=4\times 7$, $6\times 7$). This suggests a commensuration effect, although the number of plaquettes is too small to determine it. 

\section{Smaller particle numbers}\label{app:npart}
We have investigated the same plaquettes as described above filled with 4 or 5 particles. When the number of particles is changed, different Wigner crystals are allowed by the boundary conditions. However, they still follow, to large extent, the behavior of classical particles.

\begin{figure}
\includegraphics[width=0.9\textwidth]{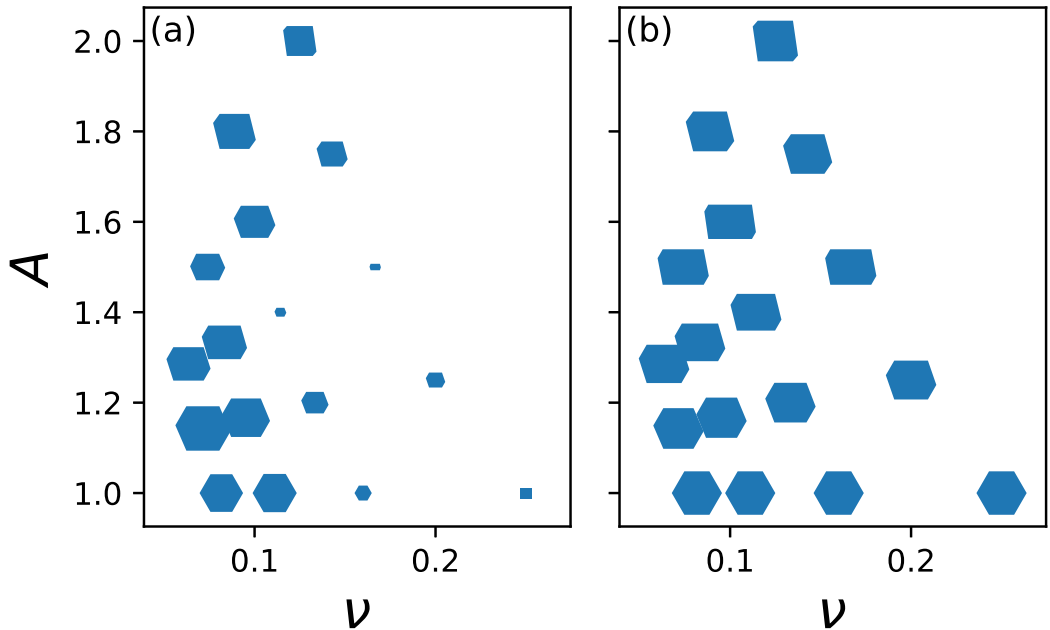}
\caption{(a) Exact-diagonalization phase diagram and (b) the classical prediction for $\Npart=4$ with $V^{\SC}_{0.5}$. }
\label{fig:4part}
\end{figure}

Figure \ref{fig:4part} shows the phase diagram for kagome lattice with $\Npart=4$ and $V^{\SC}_{0.5}$ along with the classical predictions. Note that the $L_1=L_2$ plaquettes are now included, because they do not yield degenerate Wigner crystals. The Wigner-Seitz cells of the WCs tend to be close to hexagonal for low aspect ratio (with a perfect hexagon for aspect ratio 1), while for higher aspect ratio they deviate from this shape. The agreement between classical and ED results is good. We have investigated $\Npart=4$ on kagome and honeycomb lattice with following interaction parameters: $V^{\SC}$ with $\alpha\in [0,0.6]$, and $V^{\mathrm{Log}}$ with $\beta \in [1.4,3,0]$, with both parameters varying by 0.1. Both lattices yield similar results. For every kind of interactions, the lower half of the phase diagram is similar to the one in Fig. \ref{fig:4part}. The variations in the shape of the WC exists only in the upper half of the diagram and are stronger for $V^{\mathrm{Log}}$ than for $V^{\SC}$. The shapes of the WCs agree well with the classical model, provided that the interaction is sufficiently long-range so that it does not yield degenerate ground states. It is perfect or nearly perfect (at most one plaquette predicted wrong) for logarithmic interaction, and slightly worse for the screened Coulomb potential, where typically there are two or three plaquettes where the predicted shape was different from the one in ED.

On the checkerboard lattice, we do not encounter any Wigner patterns, but the deformations of WCs are present.  Again, we try to parameterize them using a single parameter and include in the classical model. However, the predictions obtained in such a way do not reproduce the ED results. Moreover, we again note that there are two types of deformations which tend to occur mostly when $N_1$ is even and $N_2$ is odd, and vice versa. This strengthens our suggestion that this is a commensuration effect, and at least some deformations are due to the presence of lattice. If the deformations are not considered (i.e. they are not included in classical model and are regarded as regular WCs when analyzing the ED results), the classical model gives a good description of WC shapes, with perfect agreement for $V^{\mathrm{Log}}_{\beta \geq 2.1}$

For kagome and honeycomb plaquettes with $\Npart=5$ particles, the shape of Wigner crystal is the same regardless of interaction parameters in the whole range we investigated ($\alpha\in [0,1]$, $\beta \in [1.4, 3]$, changing by 0.2) and is predicted by the classical model with 100\% accuracy. What is interesting is also the disappearance of Wigner crystals at higher aspect ratios $A$ for $V^{\mathrm{Log}}_{\beta \geq 2.0}$. The WCs are not replaced by Wigner patterns, but rather by stripe-like PCD patterns. 5 particles on checkerboard lattice are much more difficult to analyze, as every possible Wigner crystal is two-fold degenerate due to reflection symmetry. Indeed, for some plaquettes and some interaction parameters we observe a PCD which can be interpreted as two such WCs superimposed. Also, we find PCDs which may be a superposition of degenerate WPs or deformed WCs. Due to the degeneracies, we decide to exclude the 5-particle checkerboard cases from our analysis.

\section{Finite-size effects -- details}\label{app:finitesize}

\begin{figure}
\includegraphics[width=0.9\textwidth]{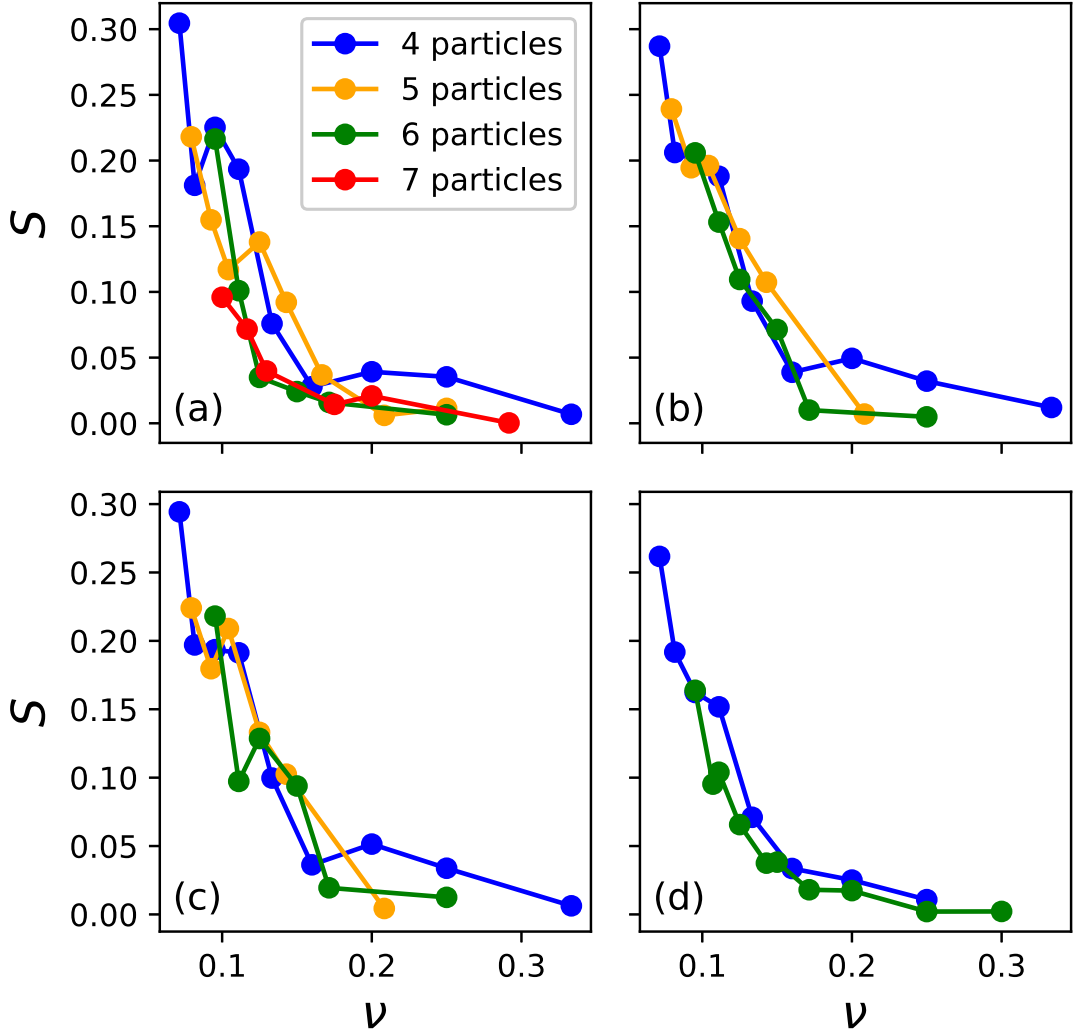}
\caption{Dependence of the crystallization strength on the filling factor for different particle number for (a)
kagome plaquettes with $V^{\mathrm{SC}}_{0.0}$, (b) honeycomb lattice plaquettes with $V^{\mathrm{SC}}_{0.5}$, (b) honeycomb lattice plaquettes with $V^{\mathrm{SC}}_{0.0}$, (d) checkerboard plaquettes with $V^{\mathrm{SC}}_{0.0}$. The aspect ratios of the plaquettes included in (a), (b), (c) varies in the following ranges: $A\in [1.0, 1.2]$ for $\Npart=4$, $A\in [1.2, 1.6]$ for $\Npart=5$,  $A\in [1.4, 1.6]$ for $\Npart=6$, $A \in [1.4, 1.67]$ for $\Npart=7$. In (d), the ranges are $A\in [1.0, 1.2]$ for $\Npart=4$, $A\in [1.14, 1.6]$ for $\Npart=6$.}
\label{fig:compare_sizes}
\end{figure}

To gain some insight on the finite size effects, we compare the results for different particle numbers. Figure \ref{fig:compare_sizes} shows comparison of the crystallization strength vs filling factor plots for four cases: (a) kagome lattice with short-range interaction $V^{\mathrm{SC}}_{0.5}$, (b) the honeycomb lattice with long-range interaction $V^{\mathrm{SC}}_{0.0}$, (b) the honeycomb lattice with long-range interaction $V^{\mathrm{SC}}_{0.0}$, (d) the checkerboard lattice with long-range interaction $V^{\mathrm{SC}}_{0.0}$. The results in all the subfigures of this Figure involve the results for $\Npart=4,5,6$, described in the previous Appendices and in the main text. Additionally, for kagome lattice we performed the calculation with $\Npart=7$, whose results are included in Fig \ref{fig:compare_sizes} (a). Also, we note that in Fig. \ref{fig:compare_sizes} (d) we plot only two curves, as the $\Npart=5$ case leads to degeneracy on the checkerboard lattice, and that for this lattice we study only the long-range interaction, as the short-range one leads to the presence of Wigner patterns at $\Npart=6$.

The results shown in all four subfigures of Fig. \ref{fig:compare_sizes} subfigures show an agreement between the crystallization strengths obtained for different particle numbers. This agreement is better for checkerboard (Fig. \ref{fig:compare_sizes} (d)) and honeycomb (Fig. \ref{fig:compare_sizes} (b) and (c)) lattices than for kagome lattice (Fig. \ref{fig:compare_sizes}). We do not observe the transition getting more sharp as the system size increases. However, as noted in the main text the extrapolation to the thermodynamic limit cannot be performed reliably, especially when the result depend strongly on sample geometry. 

\begin{figure}
\includegraphics[width=0.9\textwidth]{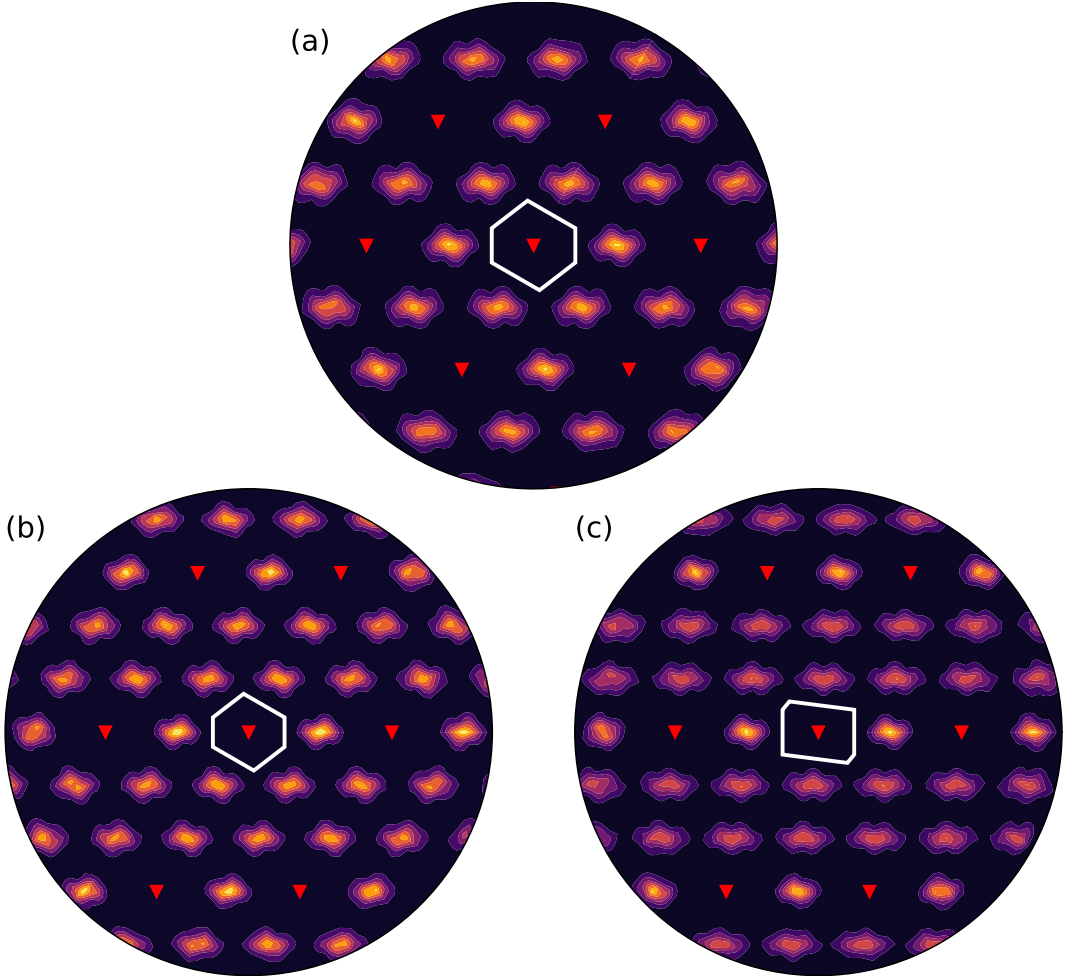}
\caption{The influence of finite size effect on the shape of the Wigner crystal. In (a), we show the pair correlation function for a $7 \times 6$ kagome plaquette with $\Npart=4$ and short-range interaction $V^{\mathrm{SC}}_{0.5}$. A PCD indistinguishable from the one shown in (a) is also obtained for long-range interaction $V^{\mathrm{SC}}_{0.0}$.  In (b), we show the PCD for a $7 \times 9$ kagome plaquette with $\Npart=6$ and $V^{\mathrm{SC}}_{0.5}$. The white shapes denote the Wigner-Seitz cells of the Wigner crystals. It can be seen that the shape of this cell is the same in (a) and (b). In (c), we show the PCD for a $7 \times 9$ kagome plaquette with $\Npart=6$ and $V^{\mathrm{SC}}_{0.5}$. Now, the Wigner-Seitz cell has a different shape than in (a).
}
\label{fig:finitesize}
\end{figure}

The finite-size effects influence also the shape of the Wigner crystal. It is difficult to investigate this effect systematically, as the boundary conditions rarely allow the formation of the Wigner crystals with the same shape and with different $\Npart$. We have such a possibility only on three pairs of plaquettes: (i) $7\times 6$ with $\Npart=4$ and $7\times 9$ with $\Npart=6$, (ii) $6\times 6$ with $\Npart=4$ and $6\times 9$ with $\Npart=6$, (iii) $5\times 6$ with $\Npart=4$ and $5\times 9$ with $\Npart=6$. Figure \ref{fig:finitesize} shows the results for pair (i) for kagome lattice and $V^{SC}$ interaction. In Fig \ref{fig:finitesize} (a) we plot the pair correlation density for a $6\times 7$ plaquette with $\Npart=4$ with $V^{SC}_{0.5}$ interaction. The white shape is the Wigner-Seitz cell of the Wigner crystal. This result can be compared with Fig \ref{fig:finitesize} (b), which shows the pair correlation density for the $7\times 9$ plaquette with $\Npart=6$. The unit cell of the Wigner crystal is the same as in Fig. \ref{fig:finitesize}  (a), suggesting that the finite-size effects do not influence the shape of the Wigner crystal. The situation becomes different when we consider the unscreened Coulomb interaction. In such a case, for the  $6\times 7$ plaquette with $\Npart=4$ we obtain a PCD pattern indistinguishable from the one in Fig \ref{fig:finitesize} (a). However, for the $7\times 9$ plaquette with $\Npart=6$, we obtain the PCD shown in Fig \ref{fig:finitesize} (c), different from the one in Fig \ref{fig:finitesize} (b). Thus, for the long-range interaction the finite-size effects are stronger and influence the shape of the WC.

A similar behavior is seen for pair (ii): the same unit cell of WC is obtained on both plaquettes if the interaction range is short, but for the long range they become different. On the other hand, for pair (iii) we get different WCs on the two plaquettes for both short and long range interaction. That is, even for the short-range interaction the finite-size effects can influence the shape of the Wigner crystal. Similar results are obtained also for honeycomb lattice. Thus, we conclude that on kagome and honeycomb lattices we observe strong finite-size effects for long range interaction and moderate finite-size effects for short-range one. 

For checkerboard lattice, we obtain strong finite-size effecs for both short- and long- range interaction. For short-range interaction, the Wigner crystals is the same on the two plaquettes only in pair (iii). On the two other pairs, the larger plaquette usually contains the Wigner pattern, which cannot be present on the smaller one. For long-range interaction, in all three pairs we obtain a rectangular WC on the smaller plaquettes and a more hexagonal one at the larger plaquettes. The only exception is the logarithmic interaction $V^{\mathrm{Log}}_{1.8}$, for which we get a WC with the same unit cell on each pair of plaquettes. 

Thus, we conclude that our calculation is prone to finite-size effects. We cannot perform a reliable extrapolation to thermodynamic limit neither for the crystallization strength nor for the crystal shape. We note that the finite-size effects related to WC shape are present also on the classical level -- for example, while in the $\Npart=6$ case the classical model predicts a non-hexagonal WC for unscreened Coulomb interaction, the infinite-plane classical Wigner crystal is hexagonal \cite{bonsall1977some}. On the other hand, as the similarity between the classical and quantum results exists for all the system sizes we investigated regardless of the number of particles and interaction types, it is possible that it will hold also in the thermodynamic limit. This does not have to be the case, as it may occur that, for example, the lattice effects will be more visible as $\Npart$ increases. However, if it is, and if Wigner crystal exists in the thermodynamic limit, we can expect that it will be hexagonal for both screened and unscreened Coulomb interaction basing for the infinite-plane results \cite{bonsall1977some, peeters1987wigner}.

\section{Comparison with trivial system}\label{app:triv}
There are two reasons to suspect that the topological properties of the flat bands may affect the Wigner crystallization. The first is the possible occurrence of FCIs on these lattices. In the course of our analysis, several plaquettes allowed for the occurence of the Laughlin fillings $\nu=1/5$ or $\nu=1/7$. At some of them, for certain interaction parameters, the lowest energy states obey the FCI counting rules \cite{BernevigCounting}. Nevertheless, for most of them the pair correlation density is not uniform, it is either WC, WP, a stripe pattern or a different, but non-uniform charge ordering. The only cases in which we are not able to disprove the presence of an FCI by looking at the pair correlation density are kagome $5\times 5$ plaquettes with 5 particles ($\nu=1/5$) for some values of interaction parameters.  However, this plaquette allows for a degenerate WC and hence is excluded from our analyses. Moreover, even if this state is an FCI, it is not a stable one, as we do not observe it for similar interaction on other $\nu=1/5$ plaquettes. Therefore, we can neglect the presence of FCIs in our analysis. 


The second reason are the constraints on particle localization forced by nontrivial topology. It is impossible to localize the Wannier function in both dimensions if the Chern number is nonzero \cite{ThoulessLocalization}. Therefore, it may mean that the Wigner crystallization in the trivial lattice would be stronger. To check this hypothesis, we have performed the calculation for trivial honeycomb system with nonzero staggered potential $\epsilon=0.15$. We have chosen four data points representing a shorter- and longer- range version of both interactions: $V^{\SC}_{0}$, $V^{\SC}_{0.5}$, $V^{\mathrm{Log}}_{1.3}$, $V^{\mathrm{Log}}_{3.0}$. Comparing the phase diagrams with the ones of nontrivial honeycomb lattice discussed previously, we discover that the shapes of Wigner crystals are exactly the same, and there were only minor changes in the crystallization strength. Therefore, we conclude that the Chern number of the flat band has no significant effect on the Wigner crystallization. We note that this is not the effect of the band mixing due to strong interaction, as all the results are obtained using the band-projected exact diagonalization.

\begin{acknowledgments} 
The authors acknowledge partial financial support from National Science Center (NCN), Poland, grant Maestro No. 2014/14/A/ST3/00654. Our calculations were performed in the Wroc\l{}aw Center for Networking and Supercomputing. 
\end{acknowledgments}
\bibliography{wigner}

\begin{thebibliography}{83}
\expandafter\ifx\csname natexlab\endcsname\relax\def\natexlab#1{#1}\fi
\expandafter\ifx\csname bibnamefont\endcsname\relax
  \def\bibnamefont#1{#1}\fi
\expandafter\ifx\csname bibfnamefont\endcsname\relax
  \def\bibfnamefont#1{#1}\fi
\expandafter\ifx\csname citenamefont\endcsname\relax
  \def\citenamefont#1{#1}\fi
\expandafter\ifx\csname url\endcsname\relax
  \def\url#1{\texttt{#1}}\fi
\expandafter\ifx\csname urlprefix\endcsname\relax\def\urlprefix{URL }\fi
\providecommand{\bibinfo}[2]{#2}
\providecommand{\eprint}[2][]{\url{#2}}

\bibitem[{\citenamefont{Haldane}(1988)}]{Haldane}
\bibinfo{author}{\bibfnamefont{F.~D.~M.} \bibnamefont{Haldane}},
  \bibinfo{journal}{Physical Review Letters} \textbf{\bibinfo{volume}{61}},
  \bibinfo{pages}{2015} (\bibinfo{year}{1988}),
  \urlprefix\url{http://link.aps.org/doi/10.1103/PhysRevLett.61.2015}.

\bibitem[{\citenamefont{Thouless et~al.}(1982)\citenamefont{Thouless, Kohmoto,
  Nightingale, and Den~Nijs}}]{TKNN}
\bibinfo{author}{\bibfnamefont{D.~J.} \bibnamefont{Thouless}},
  \bibinfo{author}{\bibfnamefont{M.}~\bibnamefont{Kohmoto}},
  \bibinfo{author}{\bibfnamefont{M.~P.} \bibnamefont{Nightingale}},
  \bibnamefont{and} \bibinfo{author}{\bibfnamefont{M.}~\bibnamefont{Den~Nijs}},
  \bibinfo{journal}{Physical Review Letters} \textbf{\bibinfo{volume}{49}},
  \bibinfo{pages}{405} (\bibinfo{year}{1982}).

\bibitem[{\citenamefont{Cooper and Wilkin}(1999)}]{cooper1999composite}
\bibinfo{author}{\bibfnamefont{N.~R.} \bibnamefont{Cooper}} \bibnamefont{and}
  \bibinfo{author}{\bibfnamefont{N.~K.} \bibnamefont{Wilkin}},
  \bibinfo{journal}{Phys. Rev. B} \textbf{\bibinfo{volume}{60}},
  \bibinfo{pages}{R16279} (\bibinfo{year}{1999}),
  \urlprefix\url{https://link.aps.org/doi/10.1103/PhysRevB.60.R16279}.

\bibitem[{\citenamefont{Jaksch and Zoller}(2003)}]{jaksch2003creation}
\bibinfo{author}{\bibfnamefont{D.}~\bibnamefont{Jaksch}} \bibnamefont{and}
  \bibinfo{author}{\bibfnamefont{P.}~\bibnamefont{Zoller}},
  \bibinfo{journal}{New Journal of Physics} \textbf{\bibinfo{volume}{5}},
  \bibinfo{pages}{56} (\bibinfo{year}{2003}),
  \urlprefix\url{http://stacks.iop.org/1367-2630/5/i=1/a=356}.

\bibitem[{\citenamefont{Aidelsburger et~al.}(2011)\citenamefont{Aidelsburger,
  Atala, Nascimb\`ene, Trotzky, Chen, and
  Bloch}}]{aidelsburger2011experimental}
\bibinfo{author}{\bibfnamefont{M.}~\bibnamefont{Aidelsburger}},
  \bibinfo{author}{\bibfnamefont{M.}~\bibnamefont{Atala}},
  \bibinfo{author}{\bibfnamefont{S.}~\bibnamefont{Nascimb\`ene}},
  \bibinfo{author}{\bibfnamefont{S.}~\bibnamefont{Trotzky}},
  \bibinfo{author}{\bibfnamefont{Y.-A.} \bibnamefont{Chen}}, \bibnamefont{and}
  \bibinfo{author}{\bibfnamefont{I.}~\bibnamefont{Bloch}},
  \bibinfo{journal}{Phys. Rev. Lett.} \textbf{\bibinfo{volume}{107}},
  \bibinfo{pages}{255301} (\bibinfo{year}{2011}),
  \urlprefix\url{https://link.aps.org/doi/10.1103/PhysRevLett.107.255301}.

\bibitem[{\citenamefont{Miyake et~al.}(2013)\citenamefont{Miyake, Siviloglou,
  Kennedy, Burton, and Ketterle}}]{miyake2013realizing}
\bibinfo{author}{\bibfnamefont{H.}~\bibnamefont{Miyake}},
  \bibinfo{author}{\bibfnamefont{G.~A.} \bibnamefont{Siviloglou}},
  \bibinfo{author}{\bibfnamefont{C.~J.} \bibnamefont{Kennedy}},
  \bibinfo{author}{\bibfnamefont{W.~C.} \bibnamefont{Burton}},
  \bibnamefont{and} \bibinfo{author}{\bibfnamefont{W.}~\bibnamefont{Ketterle}},
  \bibinfo{journal}{Phys. Rev. Lett.} \textbf{\bibinfo{volume}{111}},
  \bibinfo{pages}{185302} (\bibinfo{year}{2013}),
  \urlprefix\url{https://link.aps.org/doi/10.1103/PhysRevLett.111.185302}.

\bibitem[{\citenamefont{Aidelsburger et~al.}(2013)\citenamefont{Aidelsburger,
  Atala, Lohse, Barreiro, Paredes, and Bloch}}]{aidelsburger2013realization}
\bibinfo{author}{\bibfnamefont{M.}~\bibnamefont{Aidelsburger}},
  \bibinfo{author}{\bibfnamefont{M.}~\bibnamefont{Atala}},
  \bibinfo{author}{\bibfnamefont{M.}~\bibnamefont{Lohse}},
  \bibinfo{author}{\bibfnamefont{J.~T.} \bibnamefont{Barreiro}},
  \bibinfo{author}{\bibfnamefont{B.}~\bibnamefont{Paredes}}, \bibnamefont{and}
  \bibinfo{author}{\bibfnamefont{I.}~\bibnamefont{Bloch}},
  \bibinfo{journal}{Phys. Rev. Lett.} \textbf{\bibinfo{volume}{111}},
  \bibinfo{pages}{185301} (\bibinfo{year}{2013}),
  \urlprefix\url{https://link.aps.org/doi/10.1103/PhysRevLett.111.185301}.

\bibitem[{\citenamefont{Jotzu et~al.}(2014)\citenamefont{Jotzu, Messer,
  Desbuquois, Lebrat, Uehlinger, Greif, and Esslinger}}]{ChernExperiment2}
\bibinfo{author}{\bibfnamefont{G.}~\bibnamefont{Jotzu}},
  \bibinfo{author}{\bibfnamefont{M.}~\bibnamefont{Messer}},
  \bibinfo{author}{\bibfnamefont{R.}~\bibnamefont{Desbuquois}},
  \bibinfo{author}{\bibfnamefont{M.}~\bibnamefont{Lebrat}},
  \bibinfo{author}{\bibfnamefont{T.}~\bibnamefont{Uehlinger}},
  \bibinfo{author}{\bibfnamefont{D.}~\bibnamefont{Greif}}, \bibnamefont{and}
  \bibinfo{author}{\bibfnamefont{T.}~\bibnamefont{Esslinger}},
  \bibinfo{journal}{Nature} \textbf{\bibinfo{volume}{515}},
  \bibinfo{pages}{237} (\bibinfo{year}{2014}), ISSN \bibinfo{issn}{0028-0836},
  \bibinfo{note}{letter},
  \urlprefix\url{http://dx.doi.org/10.1038/nature13915}.

\bibitem[{\citenamefont{Chang et~al.}(2013)\citenamefont{Chang, Zhang, Feng,
  Shen, Zhang, Guo, Li, Ou, Wei, Wang et~al.}}]{ChernExperiment}
\bibinfo{author}{\bibfnamefont{C.-Z.} \bibnamefont{Chang}},
  \bibinfo{author}{\bibfnamefont{J.}~\bibnamefont{Zhang}},
  \bibinfo{author}{\bibfnamefont{X.}~\bibnamefont{Feng}},
  \bibinfo{author}{\bibfnamefont{J.}~\bibnamefont{Shen}},
  \bibinfo{author}{\bibfnamefont{Z.}~\bibnamefont{Zhang}},
  \bibinfo{author}{\bibfnamefont{M.}~\bibnamefont{Guo}},
  \bibinfo{author}{\bibfnamefont{K.}~\bibnamefont{Li}},
  \bibinfo{author}{\bibfnamefont{Y.}~\bibnamefont{Ou}},
  \bibinfo{author}{\bibfnamefont{P.}~\bibnamefont{Wei}},
  \bibinfo{author}{\bibfnamefont{L.-L.} \bibnamefont{Wang}},
  \bibnamefont{et~al.}, \bibinfo{journal}{Science}
  \textbf{\bibinfo{volume}{340}}, \bibinfo{pages}{167} (\bibinfo{year}{2013}),
  \urlprefix\url{http://science.sciencemag.org/content/340/6129/167}.

\bibitem[{\citenamefont{Liu et~al.}(2012)\citenamefont{Liu, Chen, Wang, and
  Gong}}]{flatkagome}
\bibinfo{author}{\bibfnamefont{R.}~\bibnamefont{Liu}},
  \bibinfo{author}{\bibfnamefont{W.-C.} \bibnamefont{Chen}},
  \bibinfo{author}{\bibfnamefont{Y.-F.} \bibnamefont{Wang}}, \bibnamefont{and}
  \bibinfo{author}{\bibfnamefont{C.-D.} \bibnamefont{Gong}},
  \bibinfo{journal}{Journal of Physics: Condensed Matter}
  \textbf{\bibinfo{volume}{24}}, \bibinfo{pages}{305602}
  (\bibinfo{year}{2012}),
  \urlprefix\url{http://stacks.iop.org/0953-8984/24/i=30/a=305602}.

\bibitem[{\citenamefont{Sun et~al.}(2011)\citenamefont{Sun, Gu, Katsura, and
  Das~Sarma}}]{Sun}
\bibinfo{author}{\bibfnamefont{K.}~\bibnamefont{Sun}},
  \bibinfo{author}{\bibfnamefont{Z.}~\bibnamefont{Gu}},
  \bibinfo{author}{\bibfnamefont{H.}~\bibnamefont{Katsura}}, \bibnamefont{and}
  \bibinfo{author}{\bibfnamefont{S.}~\bibnamefont{Das~Sarma}},
  \bibinfo{journal}{Physical Review Letters} \textbf{\bibinfo{volume}{106}},
  \bibinfo{pages}{236803} (\bibinfo{year}{2011}),
  \urlprefix\url{http://link.aps.org/doi/10.1103/PhysRevLett.106.236803}.

\bibitem[{\citenamefont{Tang et~al.}(2011)\citenamefont{Tang, Mei, and
  Wen}}]{Tang}
\bibinfo{author}{\bibfnamefont{E.}~\bibnamefont{Tang}},
  \bibinfo{author}{\bibfnamefont{J.-W.} \bibnamefont{Mei}}, \bibnamefont{and}
  \bibinfo{author}{\bibfnamefont{X.-G.} \bibnamefont{Wen}},
  \bibinfo{journal}{Physical Review Letters} \textbf{\bibinfo{volume}{106}},
  \bibinfo{pages}{236802} (\bibinfo{year}{2011}),
  \urlprefix\url{http://link.aps.org/doi/10.1103/PhysRevLett.106.236802}.

\bibitem[{\citenamefont{Neupert et~al.}(2011)\citenamefont{Neupert, Santos,
  Chamon, and Mudry}}]{Neupert}
\bibinfo{author}{\bibfnamefont{T.}~\bibnamefont{Neupert}},
  \bibinfo{author}{\bibfnamefont{L.}~\bibnamefont{Santos}},
  \bibinfo{author}{\bibfnamefont{C.}~\bibnamefont{Chamon}}, \bibnamefont{and}
  \bibinfo{author}{\bibfnamefont{C.}~\bibnamefont{Mudry}},
  \bibinfo{journal}{Physical Review Letters} \textbf{\bibinfo{volume}{106}},
  \bibinfo{pages}{236804} (\bibinfo{year}{2011}),
  \urlprefix\url{http://link.aps.org/doi/10.1103/PhysRevLett.106.236804}.

\bibitem[{\citenamefont{Sheng et~al.}(2011)\citenamefont{Sheng, Gu, Sun, and
  Sheng}}]{SunNature}
\bibinfo{author}{\bibfnamefont{D.}~\bibnamefont{Sheng}},
  \bibinfo{author}{\bibfnamefont{Z.-C.} \bibnamefont{Gu}, \bibfnamefont{Gu}},
  \bibinfo{author}{\bibfnamefont{K.}~\bibnamefont{Sun}}, \bibnamefont{and}
  \bibinfo{author}{\bibfnamefont{L.}~\bibnamefont{Sheng}},
  \bibinfo{journal}{Nat Commun} \textbf{\bibinfo{volume}{2}},
  \bibinfo{pages}{389} (\bibinfo{year}{2011}),
  \urlprefix\url{http://dx.doi.org/10.1038/ncomms1380}.

\bibitem[{\citenamefont{Regnault and Bernevig}(2011)}]{PRX}
\bibinfo{author}{\bibfnamefont{N.}~\bibnamefont{Regnault}} \bibnamefont{and}
  \bibinfo{author}{\bibfnamefont{B.~A.} \bibnamefont{Bernevig}},
  \bibinfo{journal}{Phys. Rev. X} \textbf{\bibinfo{volume}{1}},
  \bibinfo{pages}{021014} (\bibinfo{year}{2011}),
  \urlprefix\url{http://link.aps.org/doi/10.1103/PhysRevX.1.021014}.

\bibitem[{\citenamefont{Wu et~al.}(2012{\natexlab{a}})\citenamefont{Wu,
  Bernevig, and Regnault}}]{Zoology}
\bibinfo{author}{\bibfnamefont{Y.-L.} \bibnamefont{Wu}},
  \bibinfo{author}{\bibfnamefont{B.~A.} \bibnamefont{Bernevig}},
  \bibnamefont{and} \bibinfo{author}{\bibfnamefont{N.}~\bibnamefont{Regnault}},
  \bibinfo{journal}{Physical Review B} \textbf{\bibinfo{volume}{85}},
  \bibinfo{pages}{075116} (\bibinfo{year}{2012}{\natexlab{a}}),
  \urlprefix\url{http://link.aps.org/doi/10.1103/PhysRevB.85.075116}.

\bibitem[{\citenamefont{L\"auchli et~al.}(2013)\citenamefont{L\"auchli, Liu,
  Bergholtz, and Moessner}}]{Hierarchy}
\bibinfo{author}{\bibfnamefont{A.~M.} \bibnamefont{L\"auchli}},
  \bibinfo{author}{\bibfnamefont{Z.}~\bibnamefont{Liu}},
  \bibinfo{author}{\bibfnamefont{E.~J.} \bibnamefont{Bergholtz}},
  \bibnamefont{and} \bibinfo{author}{\bibfnamefont{R.}~\bibnamefont{Moessner}},
  \bibinfo{journal}{Physical Review Letters} \textbf{\bibinfo{volume}{111}},
  \bibinfo{pages}{126802} (\bibinfo{year}{2013}),
  \urlprefix\url{http://link.aps.org/doi/10.1103/PhysRevLett.111.126802}.

\bibitem[{\citenamefont{Liu et~al.}(2013{\natexlab{a}})\citenamefont{Liu,
  Repellin, Bernevig, and Regnault}}]{BeyondLaughlin}
\bibinfo{author}{\bibfnamefont{T.}~\bibnamefont{Liu}},
  \bibinfo{author}{\bibfnamefont{C.}~\bibnamefont{Repellin}},
  \bibinfo{author}{\bibfnamefont{B.~A.} \bibnamefont{Bernevig}},
  \bibnamefont{and} \bibinfo{author}{\bibfnamefont{N.}~\bibnamefont{Regnault}},
  \bibinfo{journal}{Physical Review B} \textbf{\bibinfo{volume}{87}},
  \bibinfo{pages}{205136} (\bibinfo{year}{2013}{\natexlab{a}}),
  \urlprefix\url{http://link.aps.org/doi/10.1103/PhysRevB.87.205136}.

\bibitem[{\citenamefont{Wang et~al.}(2012)\citenamefont{Wang, Yao, Gu, Gong,
  and Sheng}}]{MooreReadWang}
\bibinfo{author}{\bibfnamefont{Y.-F.} \bibnamefont{Wang}},
  \bibinfo{author}{\bibfnamefont{H.}~\bibnamefont{Yao}},
  \bibinfo{author}{\bibfnamefont{Z.-C.} \bibnamefont{Gu}},
  \bibinfo{author}{\bibfnamefont{C.-D.} \bibnamefont{Gong}}, \bibnamefont{and}
  \bibinfo{author}{\bibfnamefont{D.~N.} \bibnamefont{Sheng}},
  \bibinfo{journal}{Physical Review Letters} \textbf{\bibinfo{volume}{108}},
  \bibinfo{pages}{126805} (\bibinfo{year}{2012}),
  \urlprefix\url{http://link.aps.org/doi/10.1103/PhysRevLett.108.126805}.

\bibitem[{\citenamefont{Grushin et~al.}(2015)\citenamefont{Grushin, Motruk,
  Zaletel, and Pollmann}}]{Johannes}
\bibinfo{author}{\bibfnamefont{A.~G.} \bibnamefont{Grushin}},
  \bibinfo{author}{\bibfnamefont{J.}~\bibnamefont{Motruk}},
  \bibinfo{author}{\bibfnamefont{M.~P.} \bibnamefont{Zaletel}},
  \bibnamefont{and} \bibinfo{author}{\bibfnamefont{F.}~\bibnamefont{Pollmann}},
  \bibinfo{journal}{Physical Review B} \textbf{\bibinfo{volume}{91}},
  \bibinfo{pages}{035136} (\bibinfo{year}{2015}),
  \urlprefix\url{https://link.aps.org/doi/10.1103/PhysRevB.91.035136}.

\bibitem[{\citenamefont{Jaworowski et~al.}(2015)\citenamefont{Jaworowski,
  Manolescu, and Potasz}}]{We}
\bibinfo{author}{\bibfnamefont{B.}~\bibnamefont{Jaworowski}},
  \bibinfo{author}{\bibfnamefont{A.}~\bibnamefont{Manolescu}},
  \bibnamefont{and} \bibinfo{author}{\bibfnamefont{P.}~\bibnamefont{Potasz}},
  \bibinfo{journal}{Phys. Rev. B} \textbf{\bibinfo{volume}{92}},
  \bibinfo{pages}{245119} (\bibinfo{year}{2015}),
  \urlprefix\url{https://link.aps.org/doi/10.1103/PhysRevB.92.245119}.

\bibitem[{\citenamefont{Cincio and Vidal}(2013)}]{cincio2013characterizing}
\bibinfo{author}{\bibfnamefont{L.}~\bibnamefont{Cincio}} \bibnamefont{and}
  \bibinfo{author}{\bibfnamefont{G.}~\bibnamefont{Vidal}},
  \bibinfo{journal}{Physical Review Letters} \textbf{\bibinfo{volume}{110}},
  \bibinfo{pages}{067208} (\bibinfo{year}{2013}),
  \urlprefix\url{https://link.aps.org/doi/10.1103/PhysRevLett.110.067208}.

\bibitem[{\citenamefont{Liu et~al.}(2013{\natexlab{b}})\citenamefont{Liu,
  Kovrizhin, and Bergholtz}}]{liu2013bulk}
\bibinfo{author}{\bibfnamefont{Z.}~\bibnamefont{Liu}},
  \bibinfo{author}{\bibfnamefont{D.~L.} \bibnamefont{Kovrizhin}},
  \bibnamefont{and} \bibinfo{author}{\bibfnamefont{E.~J.}
  \bibnamefont{Bergholtz}}, \bibinfo{journal}{Physical Review B}
  \textbf{\bibinfo{volume}{88}}, \bibinfo{pages}{081106}
  (\bibinfo{year}{2013}{\natexlab{b}}),
  \urlprefix\url{https://link.aps.org/doi/10.1103/PhysRevB.88.081106}.

\bibitem[{\citenamefont{Scaffidi and M\"oller}(2012)}]{AdiabaticFQHE}
\bibinfo{author}{\bibfnamefont{T.}~\bibnamefont{Scaffidi}} \bibnamefont{and}
  \bibinfo{author}{\bibfnamefont{G.}~\bibnamefont{M\"oller}},
  \bibinfo{journal}{Phys. Rev. Lett.} \textbf{\bibinfo{volume}{109}},
  \bibinfo{pages}{246805} (\bibinfo{year}{2012}),
  \urlprefix\url{https://link.aps.org/doi/10.1103/PhysRevLett.109.246805}.

\bibitem[{\citenamefont{Wu et~al.}(2013)\citenamefont{Wu, Regnault, and
  Bernevig}}]{BlochPseudopotentials}
\bibinfo{author}{\bibfnamefont{Y.-L.} \bibnamefont{Wu}},
  \bibinfo{author}{\bibfnamefont{N.}~\bibnamefont{Regnault}}, \bibnamefont{and}
  \bibinfo{author}{\bibfnamefont{B.~A.} \bibnamefont{Bernevig}},
  \bibinfo{journal}{Phys. Rev. Lett.} \textbf{\bibinfo{volume}{110}},
  \bibinfo{pages}{106802} (\bibinfo{year}{2013}),
  \urlprefix\url{https://link.aps.org/doi/10.1103/PhysRevLett.110.106802}.

\bibitem[{\citenamefont{Hofstadter}(1976)}]{Hofstadter}
\bibinfo{author}{\bibfnamefont{D.~R.} \bibnamefont{Hofstadter}},
  \bibinfo{journal}{Phys. Rev. B} \textbf{\bibinfo{volume}{14}},
  \bibinfo{pages}{2239} (\bibinfo{year}{1976}),
  \urlprefix\url{https://link.aps.org/doi/10.1103/PhysRevB.14.2239}.

\bibitem[{\citenamefont{Wu et~al.}(2012{\natexlab{b}})\citenamefont{Wu, Jain,
  and Sun}}]{AdiabaticHofstadter}
\bibinfo{author}{\bibfnamefont{Y.-H.} \bibnamefont{Wu}},
  \bibinfo{author}{\bibfnamefont{J.~K.} \bibnamefont{Jain}}, \bibnamefont{and}
  \bibinfo{author}{\bibfnamefont{K.}~\bibnamefont{Sun}},
  \bibinfo{journal}{Phys. Rev. B} \textbf{\bibinfo{volume}{86}},
  \bibinfo{pages}{165129} (\bibinfo{year}{2012}{\natexlab{b}}),
  \urlprefix\url{https://link.aps.org/doi/10.1103/PhysRevB.86.165129}.

\bibitem[{\citenamefont{Andrews and M\"oller}(2018)}]{FCIDiscussion}
\bibinfo{author}{\bibfnamefont{B.}~\bibnamefont{Andrews}} \bibnamefont{and}
  \bibinfo{author}{\bibfnamefont{G.}~\bibnamefont{M\"oller}},
  \bibinfo{journal}{Phys. Rev. B} \textbf{\bibinfo{volume}{97}},
  \bibinfo{pages}{035159} (\bibinfo{year}{2018}),
  \urlprefix\url{https://link.aps.org/doi/10.1103/PhysRevB.97.035159}.

\bibitem[{\citenamefont{Spanton et~al.}(2018)\citenamefont{Spanton, Zibrov,
  Zhou, Taniguchi, Watanabe, Zaletel, and Young}}]{FCIExperiment}
\bibinfo{author}{\bibfnamefont{E.~M.} \bibnamefont{Spanton}},
  \bibinfo{author}{\bibfnamefont{A.~A.} \bibnamefont{Zibrov}},
  \bibinfo{author}{\bibfnamefont{H.}~\bibnamefont{Zhou}},
  \bibinfo{author}{\bibfnamefont{T.}~\bibnamefont{Taniguchi}},
  \bibinfo{author}{\bibfnamefont{K.}~\bibnamefont{Watanabe}},
  \bibinfo{author}{\bibfnamefont{M.~P.} \bibnamefont{Zaletel}},
  \bibnamefont{and} \bibinfo{author}{\bibfnamefont{A.~F.} \bibnamefont{Young}},
  \bibinfo{journal}{Science}  (\bibinfo{year}{2018}), ISSN
  \bibinfo{issn}{0036-8075},
  \urlprefix\url{http://science.sciencemag.org/content/early/2018/02/28/science.aan8458}.

\bibitem[{\citenamefont{S\o{}rensen et~al.}(2005)\citenamefont{S\o{}rensen,
  Demler, and Lukin}}]{sorensen2005fractional}
\bibinfo{author}{\bibfnamefont{A.~S.} \bibnamefont{S\o{}rensen}},
  \bibinfo{author}{\bibfnamefont{E.}~\bibnamefont{Demler}}, \bibnamefont{and}
  \bibinfo{author}{\bibfnamefont{M.~D.} \bibnamefont{Lukin}},
  \bibinfo{journal}{Phys. Rev. Lett.} \textbf{\bibinfo{volume}{94}},
  \bibinfo{pages}{086803} (\bibinfo{year}{2005}),
  \urlprefix\url{https://link.aps.org/doi/10.1103/PhysRevLett.94.086803}.

\bibitem[{\citenamefont{Palmer and Jaksch}(2006)}]{palmer2006high}
\bibinfo{author}{\bibfnamefont{R.~N.} \bibnamefont{Palmer}} \bibnamefont{and}
  \bibinfo{author}{\bibfnamefont{D.}~\bibnamefont{Jaksch}},
  \bibinfo{journal}{Phys. Rev. Lett.} \textbf{\bibinfo{volume}{96}},
  \bibinfo{pages}{180407} (\bibinfo{year}{2006}),
  \urlprefix\url{https://link.aps.org/doi/10.1103/PhysRevLett.96.180407}.

\bibitem[{\citenamefont{Palmer et~al.}(2008)\citenamefont{Palmer, Klein, and
  Jaksch}}]{palmer2008optical}
\bibinfo{author}{\bibfnamefont{R.~N.} \bibnamefont{Palmer}},
  \bibinfo{author}{\bibfnamefont{A.}~\bibnamefont{Klein}}, \bibnamefont{and}
  \bibinfo{author}{\bibfnamefont{D.}~\bibnamefont{Jaksch}},
  \bibinfo{journal}{Phys. Rev. A} \textbf{\bibinfo{volume}{78}},
  \bibinfo{pages}{013609} (\bibinfo{year}{2008}),
  \urlprefix\url{https://link.aps.org/doi/10.1103/PhysRevA.78.013609}.

\bibitem[{\citenamefont{Hafezi et~al.}(2007)\citenamefont{Hafezi, S\o{}rensen,
  Demler, and Lukin}}]{hafezi2007fractional}
\bibinfo{author}{\bibfnamefont{M.}~\bibnamefont{Hafezi}},
  \bibinfo{author}{\bibfnamefont{A.~S.} \bibnamefont{S\o{}rensen}},
  \bibinfo{author}{\bibfnamefont{E.}~\bibnamefont{Demler}}, \bibnamefont{and}
  \bibinfo{author}{\bibfnamefont{M.~D.} \bibnamefont{Lukin}},
  \bibinfo{journal}{Phys. Rev. A} \textbf{\bibinfo{volume}{76}},
  \bibinfo{pages}{023613} (\bibinfo{year}{2007}),
  \urlprefix\url{https://link.aps.org/doi/10.1103/PhysRevA.76.023613}.

\bibitem[{\citenamefont{M\"oller and Cooper}(2009)}]{moller2009composite}
\bibinfo{author}{\bibfnamefont{G.}~\bibnamefont{M\"oller}} \bibnamefont{and}
  \bibinfo{author}{\bibfnamefont{N.~R.} \bibnamefont{Cooper}},
  \bibinfo{journal}{Phys. Rev. Lett.} \textbf{\bibinfo{volume}{103}},
  \bibinfo{pages}{105303} (\bibinfo{year}{2009}),
  \urlprefix\url{https://link.aps.org/doi/10.1103/PhysRevLett.103.105303}.

\bibitem[{\citenamefont{Kapit and Mueller}(2010)}]{kapit2010exact}
\bibinfo{author}{\bibfnamefont{E.}~\bibnamefont{Kapit}} \bibnamefont{and}
  \bibinfo{author}{\bibfnamefont{E.}~\bibnamefont{Mueller}},
  \bibinfo{journal}{Phys. Rev. Lett.} \textbf{\bibinfo{volume}{105}},
  \bibinfo{pages}{215303} (\bibinfo{year}{2010}),
  \urlprefix\url{https://link.aps.org/doi/10.1103/PhysRevLett.105.215303}.

\bibitem[{\citenamefont{Xiao et~al.}(2011)\citenamefont{Xiao, Zhu, Ran,
  Nagaosa, and Okamoto}}]{NatureTransMet}
\bibinfo{author}{\bibfnamefont{D.}~\bibnamefont{Xiao}},
  \bibinfo{author}{\bibfnamefont{W.}~\bibnamefont{Zhu}},
  \bibinfo{author}{\bibfnamefont{Y.}~\bibnamefont{Ran}},
  \bibinfo{author}{\bibfnamefont{N.}~\bibnamefont{Nagaosa}}, \bibnamefont{and}
  \bibinfo{author}{\bibfnamefont{S.}~\bibnamefont{Okamoto}},
  \bibinfo{journal}{Nat Commun} \textbf{\bibinfo{volume}{2}},
  \bibinfo{pages}{596} (\bibinfo{year}{2011}),
  \urlprefix\url{http://www.nature.com/ncomms/journal/v2/n12/suppinfo/ncomms1602_S1.html}.

\bibitem[{\citenamefont{Venderbos et~al.}(2012)\citenamefont{Venderbos,
  Kourtis, van~den Brink, and Daghofer}}]{layers1}
\bibinfo{author}{\bibfnamefont{J.~W.~F.} \bibnamefont{Venderbos}},
  \bibinfo{author}{\bibfnamefont{S.}~\bibnamefont{Kourtis}},
  \bibinfo{author}{\bibfnamefont{J.}~\bibnamefont{van~den Brink}},
  \bibnamefont{and} \bibinfo{author}{\bibfnamefont{M.}~\bibnamefont{Daghofer}},
  \bibinfo{journal}{Phys. Rev. Lett} \textbf{\bibinfo{volume}{108}},
  \bibinfo{pages}{126405} (\bibinfo{year}{2012}),
  \urlprefix\url{http://journals.aps.org/prl/abstract/10.1103/PhysRevLett.108.126405}.

\bibitem[{\citenamefont{Venderbos et~al.}(2011)\citenamefont{Venderbos,
  Daghofer, and van~den Brink}}]{layers2}
\bibinfo{author}{\bibfnamefont{J.~W.~F.} \bibnamefont{Venderbos}},
  \bibinfo{author}{\bibfnamefont{M.}~\bibnamefont{Daghofer}}, \bibnamefont{and}
  \bibinfo{author}{\bibfnamefont{J.}~\bibnamefont{van~den Brink}},
  \bibinfo{journal}{Physical Review Letters} \textbf{\bibinfo{volume}{107}},
  \bibinfo{pages}{116401} (\bibinfo{year}{2011}),
  \urlprefix\url{http://link.aps.org/doi/10.1103/PhysRevLett.107.116401}.

\bibitem[{\citenamefont{Wigner}(1934)}]{Wigner}
\bibinfo{author}{\bibfnamefont{E.}~\bibnamefont{Wigner}},
  \bibinfo{journal}{Phys. Rev.} \textbf{\bibinfo{volume}{46}},
  \bibinfo{pages}{1002} (\bibinfo{year}{1934}),
  \urlprefix\url{https://link.aps.org/doi/10.1103/PhysRev.46.1002}.

\bibitem[{\citenamefont{Yoshioka and Fukuyama}(1979)}]{YoshiokaFukuyamaHF}
\bibinfo{author}{\bibfnamefont{D.}~\bibnamefont{Yoshioka}} \bibnamefont{and}
  \bibinfo{author}{\bibfnamefont{H.}~\bibnamefont{Fukuyama}},
  \bibinfo{journal}{Journal of the Physical Society of Japan}
  \textbf{\bibinfo{volume}{47}}, \bibinfo{pages}{394} (\bibinfo{year}{1979}),
  \urlprefix\url{http://dx.doi.org/10.1143/JPSJ.47.394}.

\bibitem[{\citenamefont{Maki and Zotos}(1983)}]{MakiZotos}
\bibinfo{author}{\bibfnamefont{K.}~\bibnamefont{Maki}} \bibnamefont{and}
  \bibinfo{author}{\bibfnamefont{X.}~\bibnamefont{Zotos}},
  \bibinfo{journal}{Physical Review B} \textbf{\bibinfo{volume}{28}},
  \bibinfo{pages}{4349} (\bibinfo{year}{1983}),
  \urlprefix\url{https://link.aps.org/doi/10.1103/PhysRevB.28.4349}.

\bibitem[{\citenamefont{Maksym}(1992)}]{MaksymED}
\bibinfo{author}{\bibfnamefont{P.}~\bibnamefont{Maksym}},
  \bibinfo{journal}{Journal of Physics: Condensed Matter}
  \textbf{\bibinfo{volume}{4}}, \bibinfo{pages}{L97} (\bibinfo{year}{1992}),
  \urlprefix\url{http://stacks.iop.org/0953-8984/4/i=6/a=004}.

\bibitem[{\citenamefont{Hutchinson et~al.}(1996)\citenamefont{Hutchinson,
  Inkson, and Rowe}}]{HutchinsonED}
\bibinfo{author}{\bibfnamefont{D.}~\bibnamefont{Hutchinson}},
  \bibinfo{author}{\bibfnamefont{J.}~\bibnamefont{Inkson}}, \bibnamefont{and}
  \bibinfo{author}{\bibfnamefont{J.}~\bibnamefont{Rowe}},
  \bibinfo{journal}{Solid State Communications} \textbf{\bibinfo{volume}{97}},
  \bibinfo{pages}{515} (\bibinfo{year}{1996}),
  \urlprefix\url{http://www.sciencedirect.com/science/article/pii/0038109895006060}.

\bibitem[{\citenamefont{Yang et~al.}(2001)\citenamefont{Yang, Haldane, and
  Rezayi}}]{HaldaneED}
\bibinfo{author}{\bibfnamefont{K.}~\bibnamefont{Yang}},
  \bibinfo{author}{\bibfnamefont{F.~D.~M.} \bibnamefont{Haldane}},
  \bibnamefont{and} \bibinfo{author}{\bibfnamefont{E.~H.}
  \bibnamefont{Rezayi}}, \bibinfo{journal}{Physical Review B}
  \textbf{\bibinfo{volume}{64}}, \bibinfo{pages}{081301}
  (\bibinfo{year}{2001}),
  \urlprefix\url{https://link.aps.org/doi/10.1103/PhysRevB.64.081301}.

\bibitem[{\citenamefont{Shibata and Yoshioka}(2003)}]{ShibataYoshiokaDMRG}
\bibinfo{author}{\bibfnamefont{N.}~\bibnamefont{Shibata}} \bibnamefont{and}
  \bibinfo{author}{\bibfnamefont{D.}~\bibnamefont{Yoshioka}},
  \bibinfo{journal}{Journal of the Physical Society of Japan}
  \textbf{\bibinfo{volume}{72}}, \bibinfo{pages}{664} (\bibinfo{year}{2003}),
  \urlprefix\url{http://dx.doi.org/10.1143/JPSJ.72.664}.

\bibitem[{\citenamefont{Zhu and Louie}(1993)}]{ZhuQMC}
\bibinfo{author}{\bibfnamefont{X.}~\bibnamefont{Zhu}} \bibnamefont{and}
  \bibinfo{author}{\bibfnamefont{S.~G.} \bibnamefont{Louie}},
  \bibinfo{journal}{Physical Review Letters} \textbf{\bibinfo{volume}{70}},
  \bibinfo{pages}{335} (\bibinfo{year}{1993}),
  \urlprefix\url{https://link.aps.org/doi/10.1103/PhysRevLett.70.335}.

\bibitem[{\citenamefont{Yi and Fertig}(1998)}]{YiQMC}
\bibinfo{author}{\bibfnamefont{H.}~\bibnamefont{Yi}} \bibnamefont{and}
  \bibinfo{author}{\bibfnamefont{H.~A.} \bibnamefont{Fertig}},
  \bibinfo{journal}{Physical Review B} \textbf{\bibinfo{volume}{58}},
  \bibinfo{pages}{4019} (\bibinfo{year}{1998}),
  \urlprefix\url{https://link.aps.org/doi/10.1103/PhysRevB.58.4019}.

\bibitem[{\citenamefont{Lam and Girvin}(1984)}]{LamGirvin}
\bibinfo{author}{\bibfnamefont{P.~K.} \bibnamefont{Lam}} \bibnamefont{and}
  \bibinfo{author}{\bibfnamefont{S.~M.} \bibnamefont{Girvin}},
  \bibinfo{journal}{Physical Review B} \textbf{\bibinfo{volume}{30}},
  \bibinfo{pages}{473} (\bibinfo{year}{1984}),
  \urlprefix\url{https://link.aps.org/doi/10.1103/PhysRevB.30.473}.

\bibitem[{\citenamefont{M{\"u}ller and Koonin}(1996)}]{muller1996phase}
\bibinfo{author}{\bibfnamefont{H.-M.} \bibnamefont{M{\"u}ller}}
  \bibnamefont{and} \bibinfo{author}{\bibfnamefont{S.~E.}
  \bibnamefont{Koonin}}, \bibinfo{journal}{Physical Review B}
  \textbf{\bibinfo{volume}{54}}, \bibinfo{pages}{14532} (\bibinfo{year}{1996}),
  \urlprefix\url{https://link.aps.org/doi/10.1103/PhysRevB.54.14532}.

\bibitem[{\citenamefont{Maksym et~al.}(2000)\citenamefont{Maksym, Imamura,
  Mallon, and Aoki}}]{maksym2000molecular}
\bibinfo{author}{\bibfnamefont{P.~A.} \bibnamefont{Maksym}},
  \bibinfo{author}{\bibfnamefont{H.}~\bibnamefont{Imamura}},
  \bibinfo{author}{\bibfnamefont{G.}~\bibnamefont{Mallon}}, \bibnamefont{and}
  \bibinfo{author}{\bibfnamefont{H.}~\bibnamefont{Aoki}},
  \bibinfo{journal}{Journal of Physics: Condensed Matter}
  \textbf{\bibinfo{volume}{12}}, \bibinfo{pages}{R299} (\bibinfo{year}{2000}),
  \urlprefix\url{http://stacks.iop.org/0953-8984/12/i=22/a=201}.

\bibitem[{\citenamefont{Reimann et~al.}(2000)\citenamefont{Reimann, Koskinen,
  and Manninen}}]{reimann2000formation}
\bibinfo{author}{\bibfnamefont{S.~M.} \bibnamefont{Reimann}},
  \bibinfo{author}{\bibfnamefont{M.}~\bibnamefont{Koskinen}}, \bibnamefont{and}
  \bibinfo{author}{\bibfnamefont{M.}~\bibnamefont{Manninen}},
  \bibinfo{journal}{Physical Review B} \textbf{\bibinfo{volume}{62}},
  \bibinfo{pages}{8108} (\bibinfo{year}{2000}),
  \urlprefix\url{https://link.aps.org/doi/10.1103/PhysRevB.62.8108}.

\bibitem[{\citenamefont{Ghosal et~al.}(2007)\citenamefont{Ghosal,
  G{\"u}{\c{c}}l{\"u}, Umrigar, Ullmo, and Baranger}}]{DevrimDotWC1}
\bibinfo{author}{\bibfnamefont{A.}~\bibnamefont{Ghosal}},
  \bibinfo{author}{\bibfnamefont{A.~D.} \bibnamefont{G{\"u}{\c{c}}l{\"u}}},
  \bibinfo{author}{\bibfnamefont{C.~J.} \bibnamefont{Umrigar}},
  \bibinfo{author}{\bibfnamefont{D.}~\bibnamefont{Ullmo}}, \bibnamefont{and}
  \bibinfo{author}{\bibfnamefont{H.~U.} \bibnamefont{Baranger}},
  \bibinfo{journal}{Physical Review B} \textbf{\bibinfo{volume}{76}},
  \bibinfo{pages}{085341} (\bibinfo{year}{2007}),
  \urlprefix\url{https://link.aps.org/doi/10.1103/PhysRevB.76.085341}.

\bibitem[{\citenamefont{G{\"u}{\c{c}}l{\"u}
  et~al.}(2008)\citenamefont{G{\"u}{\c{c}}l{\"u}, Ghosal, Umrigar, and
  Baranger}}]{DevrimDotWC2}
\bibinfo{author}{\bibfnamefont{A.~D.} \bibnamefont{G{\"u}{\c{c}}l{\"u}}},
  \bibinfo{author}{\bibfnamefont{A.}~\bibnamefont{Ghosal}},
  \bibinfo{author}{\bibfnamefont{C.~J.} \bibnamefont{Umrigar}},
  \bibnamefont{and} \bibinfo{author}{\bibfnamefont{H.~U.}
  \bibnamefont{Baranger}}, \bibinfo{journal}{Physical Review B}
  \textbf{\bibinfo{volume}{77}}, \bibinfo{pages}{041301}
  (\bibinfo{year}{2008}),
  \urlprefix\url{https://link.aps.org/doi/10.1103/PhysRevB.77.041301}.

\bibitem[{\citenamefont{Grimes and Adams}(1979)}]{helium}
\bibinfo{author}{\bibfnamefont{C.~C.} \bibnamefont{Grimes}} \bibnamefont{and}
  \bibinfo{author}{\bibfnamefont{G.}~\bibnamefont{Adams}},
  \bibinfo{journal}{Physical Review Letters} \textbf{\bibinfo{volume}{42}},
  \bibinfo{pages}{795} (\bibinfo{year}{1979}),
  \urlprefix\url{https://link.aps.org/doi/10.1103/PhysRevLett.42.795}.

\bibitem[{\citenamefont{Hew et~al.}(2009)\citenamefont{Hew, Thomas, Pepper,
  Farrer, Anderson, Jones, and Ritchie}}]{wireWC}
\bibinfo{author}{\bibfnamefont{W.~K.} \bibnamefont{Hew}},
  \bibinfo{author}{\bibfnamefont{K.~J.} \bibnamefont{Thomas}},
  \bibinfo{author}{\bibfnamefont{M.}~\bibnamefont{Pepper}},
  \bibinfo{author}{\bibfnamefont{I.}~\bibnamefont{Farrer}},
  \bibinfo{author}{\bibfnamefont{D.}~\bibnamefont{Anderson}},
  \bibinfo{author}{\bibfnamefont{G.~A.~C.} \bibnamefont{Jones}},
  \bibnamefont{and} \bibinfo{author}{\bibfnamefont{D.~A.}
  \bibnamefont{Ritchie}}, \bibinfo{journal}{Physical review letters}
  \textbf{\bibinfo{volume}{102}}, \bibinfo{pages}{056804}
  (\bibinfo{year}{2009}),
  \urlprefix\url{https://link.aps.org/doi/10.1103/PhysRevLett.102.056804}.

\bibitem[{\citenamefont{G\"u\c{c}l\"u et~al.}(2009)\citenamefont{G\"u\c{c}l\"u,
  Umrigar, Jiang, and Baranger}}]{DevrimWireWC}
\bibinfo{author}{\bibfnamefont{A.~D.} \bibnamefont{G\"u\c{c}l\"u}},
  \bibinfo{author}{\bibfnamefont{C.~J.} \bibnamefont{Umrigar}},
  \bibinfo{author}{\bibfnamefont{H.}~\bibnamefont{Jiang}}, \bibnamefont{and}
  \bibinfo{author}{\bibfnamefont{H.~U.} \bibnamefont{Baranger}},
  \bibinfo{journal}{Physical Review B} \textbf{\bibinfo{volume}{80}},
  \bibinfo{pages}{201302} (\bibinfo{year}{2009}),
  \urlprefix\url{https://link.aps.org/doi/10.1103/PhysRevB.80.201302}.

\bibitem[{\citenamefont{Ziani et~al.}(2015)\citenamefont{Ziani, Cr\'epin, and
  Trauzettel}}]{ziani2015fractional}
\bibinfo{author}{\bibfnamefont{N.~T.} \bibnamefont{Ziani}},
  \bibinfo{author}{\bibfnamefont{F.}~\bibnamefont{Cr\'epin}}, \bibnamefont{and}
  \bibinfo{author}{\bibfnamefont{B.}~\bibnamefont{Trauzettel}},
  \bibinfo{journal}{Physical Review Letters} \textbf{\bibinfo{volume}{115}},
  \bibinfo{pages}{206402} (\bibinfo{year}{2015}),
  \urlprefix\url{https://link.aps.org/doi/10.1103/PhysRevLett.115.206402}.

\bibitem[{\citenamefont{De~Beule et~al.}(2016)\citenamefont{De~Beule, Ziani,
  Zarenia, Partoens, and Trauzettel}}]{beule2016correlation}
\bibinfo{author}{\bibfnamefont{C.}~\bibnamefont{De~Beule}},
  \bibinfo{author}{\bibfnamefont{N.~T.} \bibnamefont{Ziani}},
  \bibinfo{author}{\bibfnamefont{M.}~\bibnamefont{Zarenia}},
  \bibinfo{author}{\bibfnamefont{B.}~\bibnamefont{Partoens}}, \bibnamefont{and}
  \bibinfo{author}{\bibfnamefont{B.}~\bibnamefont{Trauzettel}},
  \bibinfo{journal}{Physical Review B} \textbf{\bibinfo{volume}{94}},
  \bibinfo{pages}{155111} (\bibinfo{year}{2016}),
  \urlprefix\url{https://link.aps.org/doi/10.1103/PhysRevB.94.155111}.

\bibitem[{\citenamefont{Fratini and Merino}(2009)}]{Fratini}
\bibinfo{author}{\bibfnamefont{S.}~\bibnamefont{Fratini}} \bibnamefont{and}
  \bibinfo{author}{\bibfnamefont{J.}~\bibnamefont{Merino}},
  \bibinfo{journal}{Physical Review B} \textbf{\bibinfo{volume}{80}},
  \bibinfo{pages}{165110} (\bibinfo{year}{2009}),
  \urlprefix\url{https://link.aps.org/doi/10.1103/PhysRevB.80.165110}.

\bibitem[{\citenamefont{Noda and Imada}(2002)}]{Noda}
\bibinfo{author}{\bibfnamefont{Y.}~\bibnamefont{Noda}} \bibnamefont{and}
  \bibinfo{author}{\bibfnamefont{M.}~\bibnamefont{Imada}},
  \bibinfo{journal}{Physical Review Letters} \textbf{\bibinfo{volume}{89}},
  \bibinfo{pages}{176803} (\bibinfo{year}{2002}),
  \urlprefix\url{https://link.aps.org/doi/10.1103/PhysRevLett.89.176803}.

\bibitem[{\citenamefont{Wu et~al.}(2007)\citenamefont{Wu, Bergman, Balents, and
  Das~Sarma}}]{DasSarmaWCHoneycomb}
\bibinfo{author}{\bibfnamefont{C.}~\bibnamefont{Wu}},
  \bibinfo{author}{\bibfnamefont{D.}~\bibnamefont{Bergman}},
  \bibinfo{author}{\bibfnamefont{L.}~\bibnamefont{Balents}}, \bibnamefont{and}
  \bibinfo{author}{\bibfnamefont{S.}~\bibnamefont{Das~Sarma}},
  \bibinfo{journal}{Physical Review Letters} \textbf{\bibinfo{volume}{99}},
  \bibinfo{pages}{070401} (\bibinfo{year}{2007}),
  \urlprefix\url{https://link.aps.org/doi/10.1103/PhysRevLett.99.070401}.

\bibitem[{\citenamefont{G{\"u}{\c{c}}l{\"u}}(2016)}]{DevrimGrapheneWC1}
\bibinfo{author}{\bibfnamefont{A.~D.} \bibnamefont{G{\"u}{\c{c}}l{\"u}}},
  \bibinfo{journal}{Physical Review B} \textbf{\bibinfo{volume}{93}},
  \bibinfo{pages}{045114} (\bibinfo{year}{2016}),
  \urlprefix\url{https://link.aps.org/doi/10.1103/PhysRevB.93.045114}.

\bibitem[{\citenamefont{Modarresi and
  G{\"u}{\c{c}}l{\"u}}(2017)}]{DevrimGrapheneWC2}
\bibinfo{author}{\bibfnamefont{M.}~\bibnamefont{Modarresi}} \bibnamefont{and}
  \bibinfo{author}{\bibfnamefont{A.~D.} \bibnamefont{G{\"u}{\c{c}}l{\"u}}},
  \bibinfo{journal}{Physical Review B} \textbf{\bibinfo{volume}{95}},
  \bibinfo{pages}{235103} (\bibinfo{year}{2017}),
  \urlprefix\url{https://link.aps.org/doi/10.1103/PhysRevB.95.235103}.

\bibitem[{\citenamefont{Andrei et~al.}(1988)\citenamefont{Andrei, Deville,
  Glattli, Williams, Paris, and Etienne}}]{Andrei}
\bibinfo{author}{\bibfnamefont{E.~Y.} \bibnamefont{Andrei}},
  \bibinfo{author}{\bibfnamefont{G.}~\bibnamefont{Deville}},
  \bibinfo{author}{\bibfnamefont{D.~C.} \bibnamefont{Glattli}},
  \bibinfo{author}{\bibfnamefont{F.~I.~B.} \bibnamefont{Williams}},
  \bibinfo{author}{\bibfnamefont{E.}~\bibnamefont{Paris}}, \bibnamefont{and}
  \bibinfo{author}{\bibfnamefont{B.}~\bibnamefont{Etienne}},
  \bibinfo{journal}{Physical Review Letters} \textbf{\bibinfo{volume}{60}},
  \bibinfo{pages}{2765} (\bibinfo{year}{1988}),
  \urlprefix\url{https://link.aps.org/doi/10.1103/PhysRevLett.60.2765}.

\bibitem[{\citenamefont{Kukushkin et~al.}(1994)\citenamefont{Kukushkin, Fal'ko,
  Haug, von Klitzing, Eberl, and T\"otemayer}}]{Kukushkin}
\bibinfo{author}{\bibfnamefont{I.~V.} \bibnamefont{Kukushkin}},
  \bibinfo{author}{\bibfnamefont{V.~I.} \bibnamefont{Fal'ko}},
  \bibinfo{author}{\bibfnamefont{R.~J.} \bibnamefont{Haug}},
  \bibinfo{author}{\bibfnamefont{K.}~\bibnamefont{von Klitzing}},
  \bibinfo{author}{\bibfnamefont{K.}~\bibnamefont{Eberl}}, \bibnamefont{and}
  \bibinfo{author}{\bibfnamefont{K.}~\bibnamefont{T\"otemayer}},
  \bibinfo{journal}{Physical Review Letters} \textbf{\bibinfo{volume}{72}},
  \bibinfo{pages}{3594} (\bibinfo{year}{1994}),
  \urlprefix\url{https://link.aps.org/doi/10.1103/PhysRevLett.72.3594}.

\bibitem[{\citenamefont{Haldane}(1983)}]{HaldanePseudopotentials}
\bibinfo{author}{\bibfnamefont{F.~D.~M.} \bibnamefont{Haldane}},
  \bibinfo{journal}{Physical Review Letters} \textbf{\bibinfo{volume}{51}},
  \bibinfo{pages}{605} (\bibinfo{year}{1983}),
  \urlprefix\url{https://link.aps.org/doi/10.1103/PhysRevLett.51.605}.

\bibitem[{\citenamefont{Trugman and Kivelson}(1985)}]{TrugmanKivelson}
\bibinfo{author}{\bibfnamefont{S.~A.} \bibnamefont{Trugman}} \bibnamefont{and}
  \bibinfo{author}{\bibfnamefont{S.}~\bibnamefont{Kivelson}},
  \bibinfo{journal}{Physical Review B} \textbf{\bibinfo{volume}{31}},
  \bibinfo{pages}{5280} (\bibinfo{year}{1985}),
  \urlprefix\url{https://link.aps.org/doi/10.1103/PhysRevB.31.5280}.

\bibitem[{\citenamefont{Thiebaut et~al.}(2015)\citenamefont{Thiebaut, Regnault,
  and Goerbig}}]{thiebaut2015fractional}
\bibinfo{author}{\bibfnamefont{N.}~\bibnamefont{Thiebaut}},
  \bibinfo{author}{\bibfnamefont{N.}~\bibnamefont{Regnault}}, \bibnamefont{and}
  \bibinfo{author}{\bibfnamefont{M.~O.} \bibnamefont{Goerbig}},
  \bibinfo{journal}{Physical Review B} \textbf{\bibinfo{volume}{92}},
  \bibinfo{pages}{245401} (\bibinfo{year}{2015}),
  \urlprefix\url{https://link.aps.org/doi/10.1103/PhysRevB.92.245401}.

\bibitem[{\citenamefont{Varney et~al.}(2010)\citenamefont{Varney, Sun, Rigol,
  and Galitski}}]{varney2010interaction}
\bibinfo{author}{\bibfnamefont{C.~N.} \bibnamefont{Varney}},
  \bibinfo{author}{\bibfnamefont{K.}~\bibnamefont{Sun}},
  \bibinfo{author}{\bibfnamefont{M.}~\bibnamefont{Rigol}}, \bibnamefont{and}
  \bibinfo{author}{\bibfnamefont{V.}~\bibnamefont{Galitski}},
  \bibinfo{journal}{Physical Review B} \textbf{\bibinfo{volume}{82}},
  \bibinfo{pages}{115125} (\bibinfo{year}{2010}),
  \urlprefix\url{https://link.aps.org/doi/10.1103/PhysRevB.82.115125}.

\bibitem[{\citenamefont{Grushin et~al.}(2012)\citenamefont{Grushin, Neupert,
  Chamon, and Mudry}}]{SingleParticleGrushin}
\bibinfo{author}{\bibfnamefont{A.~G.} \bibnamefont{Grushin}},
  \bibinfo{author}{\bibfnamefont{T.}~\bibnamefont{Neupert}},
  \bibinfo{author}{\bibfnamefont{C.}~\bibnamefont{Chamon}}, \bibnamefont{and}
  \bibinfo{author}{\bibfnamefont{C.}~\bibnamefont{Mudry}},
  \bibinfo{journal}{Physical Review B} \textbf{\bibinfo{volume}{86}},
  \bibinfo{pages}{205125} (\bibinfo{year}{2012}),
  \urlprefix\url{http://link.aps.org/doi/10.1103/PhysRevB.86.205125}.

\bibitem[{\citenamefont{Kourtis et~al.}(2012)\citenamefont{Kourtis, Venderbos,
  and Daghofer}}]{KourtisTriangular}
\bibinfo{author}{\bibfnamefont{S.}~\bibnamefont{Kourtis}},
  \bibinfo{author}{\bibfnamefont{J.~W.~F.} \bibnamefont{Venderbos}},
  \bibnamefont{and} \bibinfo{author}{\bibfnamefont{M.}~\bibnamefont{Daghofer}},
  \bibinfo{journal}{Physical Review B} \textbf{\bibinfo{volume}{86}},
  \bibinfo{pages}{235118} (\bibinfo{year}{2012}),
  \urlprefix\url{http://link.aps.org/doi/10.1103/PhysRevB.86.235118}.

\bibitem[{\citenamefont{Kourtis et~al.}(2017)\citenamefont{Kourtis, Neupert,
  Mudry, Sigrist, and Chen}}]{kourtis2017weyl}
\bibinfo{author}{\bibfnamefont{S.}~\bibnamefont{Kourtis}},
  \bibinfo{author}{\bibfnamefont{T.}~\bibnamefont{Neupert}},
  \bibinfo{author}{\bibfnamefont{C.}~\bibnamefont{Mudry}},
  \bibinfo{author}{\bibfnamefont{M.}~\bibnamefont{Sigrist}}, \bibnamefont{and}
  \bibinfo{author}{\bibfnamefont{W.}~\bibnamefont{Chen}},
  \bibinfo{journal}{Physical Review B} \textbf{\bibinfo{volume}{96}},
  \bibinfo{pages}{205117} (\bibinfo{year}{2017}),
  \urlprefix\url{https://link.aps.org/doi/10.1103/PhysRevB.96.205117}.

\bibitem[{\citenamefont{Kourtis and Daghofer}(2014)}]{kourtis2014combined}
\bibinfo{author}{\bibfnamefont{S.}~\bibnamefont{Kourtis}} \bibnamefont{and}
  \bibinfo{author}{\bibfnamefont{M.}~\bibnamefont{Daghofer}},
  \bibinfo{journal}{Physical Review Letters} \textbf{\bibinfo{volume}{113}},
  \bibinfo{pages}{216404} (\bibinfo{year}{2014}),
  \urlprefix\url{https://link.aps.org/doi/10.1103/PhysRevLett.113.216404}.

\bibitem[{\citenamefont{Li et~al.}(2014)\citenamefont{Li, Liu, Wu, and
  Chen}}]{LiFCIWC}
\bibinfo{author}{\bibfnamefont{W.}~\bibnamefont{Li}},
  \bibinfo{author}{\bibfnamefont{Z.}~\bibnamefont{Liu}},
  \bibinfo{author}{\bibfnamefont{Y.-S.} \bibnamefont{Wu}}, \bibnamefont{and}
  \bibinfo{author}{\bibfnamefont{Y.}~\bibnamefont{Chen}},
  \bibinfo{journal}{Physical Review B} \textbf{\bibinfo{volume}{89}},
  \bibinfo{pages}{125411} (\bibinfo{year}{2014}),
  \urlprefix\url{https://link.aps.org/doi/10.1103/PhysRevB.89.125411}.

\bibitem[{\citenamefont{{Kourtis}}(2017)}]{kourtis2017symmetry}
\bibinfo{author}{\bibfnamefont{S.}~\bibnamefont{{Kourtis}}},
  \bibinfo{journal}{ArXiv e-prints}  (\bibinfo{year}{2017}),
  \eprint{1711.00017}.

\bibitem[{\citenamefont{Zeng and Yin}(2015)}]{PhysRevB.91.075102}
\bibinfo{author}{\bibfnamefont{T.-S.} \bibnamefont{Zeng}} \bibnamefont{and}
  \bibinfo{author}{\bibfnamefont{L.}~\bibnamefont{Yin}},
  \bibinfo{journal}{Physical Review B} \textbf{\bibinfo{volume}{91}},
  \bibinfo{pages}{075102} (\bibinfo{year}{2015}),
  \urlprefix\url{https://link.aps.org/doi/10.1103/PhysRevB.91.075102}.

\bibitem[{\citenamefont{Bulu\c{c} et~al.}(2009)\citenamefont{Bulu\c{c},
  Fineman, Frigo, Gilbert, and Leiserson}}]{csb}
\bibinfo{author}{\bibfnamefont{A.}~\bibnamefont{Bulu\c{c}}},
  \bibinfo{author}{\bibfnamefont{J.~T.} \bibnamefont{Fineman}},
  \bibinfo{author}{\bibfnamefont{M.}~\bibnamefont{Frigo}},
  \bibinfo{author}{\bibfnamefont{J.~R.} \bibnamefont{Gilbert}},
  \bibnamefont{and} \bibinfo{author}{\bibfnamefont{C.~E.}
  \bibnamefont{Leiserson}}, in \emph{\bibinfo{booktitle}{Proceedings of the
  Twenty-first Annual Symposium on Parallelism in Algorithms and
  Architectures}} (\bibinfo{publisher}{ACM}, \bibinfo{address}{New York, NY,
  USA}, \bibinfo{year}{2009}), SPAA '09, pp. \bibinfo{pages}{233--244}, ISBN
  \bibinfo{isbn}{978-1-60558-606-9},
  \urlprefix\url{http://doi.acm.org/10.1145/1583991.1584053}.

\bibitem[{\citenamefont{Reichhardt and Reichhardt}(2007)}]{vortices1}
\bibinfo{author}{\bibfnamefont{C.}~\bibnamefont{Reichhardt}} \bibnamefont{and}
  \bibinfo{author}{\bibfnamefont{C.~J.~O.} \bibnamefont{Reichhardt}},
  \bibinfo{journal}{Phys. Rev. B} \textbf{\bibinfo{volume}{76}},
  \bibinfo{pages}{064523} (\bibinfo{year}{2007}),
  \urlprefix\url{https://link.aps.org/doi/10.1103/PhysRevB.76.064523}.

\bibitem[{\citenamefont{Reichhardt and Olson~Reichhardt}(2010)}]{vortices2}
\bibinfo{author}{\bibfnamefont{C.}~\bibnamefont{Reichhardt}} \bibnamefont{and}
  \bibinfo{author}{\bibfnamefont{C.~J.} \bibnamefont{Olson~Reichhardt}},
  \bibinfo{journal}{Phys. Rev. B} \textbf{\bibinfo{volume}{81}},
  \bibinfo{pages}{024510} (\bibinfo{year}{2010}),
  \urlprefix\url{https://link.aps.org/doi/10.1103/PhysRevB.81.024510}.

\bibitem[{\citenamefont{Bonsall and Maradudin}(1977)}]{bonsall1977some}
\bibinfo{author}{\bibfnamefont{L.}~\bibnamefont{Bonsall}} \bibnamefont{and}
  \bibinfo{author}{\bibfnamefont{A.~A.} \bibnamefont{Maradudin}},
  \bibinfo{journal}{Phys. Rev. B} \textbf{\bibinfo{volume}{15}},
  \bibinfo{pages}{1959} (\bibinfo{year}{1977}),
  \urlprefix\url{https://link.aps.org/doi/10.1103/PhysRevB.15.1959}.

\bibitem[{\citenamefont{Peeters and Wu}(1987)}]{peeters1987wigner}
\bibinfo{author}{\bibfnamefont{F.~M.} \bibnamefont{Peeters}} \bibnamefont{and}
  \bibinfo{author}{\bibfnamefont{X.}~\bibnamefont{Wu}}, \bibinfo{journal}{Phys.
  Rev. A} \textbf{\bibinfo{volume}{35}}, \bibinfo{pages}{3109}
  (\bibinfo{year}{1987}),
  \urlprefix\url{https://link.aps.org/doi/10.1103/PhysRevA.35.3109}.

\bibitem[{\citenamefont{Bernevig and Regnault}(2012)}]{BernevigCounting}
\bibinfo{author}{\bibfnamefont{B.~A.} \bibnamefont{Bernevig}} \bibnamefont{and}
  \bibinfo{author}{\bibfnamefont{N.}~\bibnamefont{Regnault}},
  \bibinfo{journal}{Physical Review B} \textbf{\bibinfo{volume}{85}},
  \bibinfo{pages}{075128} (\bibinfo{year}{2012}),
  \urlprefix\url{http://link.aps.org/doi/10.1103/PhysRevB.85.075128}.

\bibitem[{\citenamefont{Thouless}(1984)}]{ThoulessLocalization}
\bibinfo{author}{\bibfnamefont{D.~J.} \bibnamefont{Thouless}},
  \bibinfo{journal}{Journal of Physics C: Solid State Physics}
  \textbf{\bibinfo{volume}{17}}, \bibinfo{pages}{L325} (\bibinfo{year}{1984}),
  \urlprefix\url{http://stacks.iop.org/0022-3719/17/i=12/a=003}.

\end{thebibliography}
\end{document}